\documentclass[aps,prc,floatfix]{revtex4}

\usepackage{epsfig}
\usepackage{amsmath}

\begin{document}

\title{Intensity interferometry of thermal photons from relativistic
heavy ion collisions}
\author{Dinesh Kumar Srivastava}
\affiliation{Variable Energy Cyclotron 
Centre, 1/AF Bidhan Nagar, Kolkata 700 064, India}            
\date{\today}

\begin{abstract}
Intensity interferometry of 
thermal photons, having transverse momenta  $k_T \approx $ 0.1 -- 2.0 GeV,
 produced in relativistic
collision of heavy nuclei is studied. It is seen to provide an accurate
information about the temporal and spatial structure of the interacting system.
The source dimensions and their $k_T$ dependence revealed by the photon
interferometry, display a richness not seen in pion 
interferometry. We attribute
this to difference in the source functions, the fact that photons come out
from every stage of the collision and from every point in the system, 
and the fact  that the rate of production of photons is 
different for the quark-gluon plasma, which dominates 
the early hot stage, and the
hadronic matter which populates the last phase of the collision dynamics.
 The usefulness of this procedure is demonstrated by an application
to collision of lead nuclei at the CERN SPS.
Prediction for the
transverse momentum dependence of the sizes for SPS, RHIC, and LHC energies
are given.
\end{abstract}

\pacs{25.75.-q,12.38.Mh}
\maketitle
\section{Introduction}
The search for quark hadron phase transition and the investigation of
 the properties of quark gluon plasma stand among the most challenging 
as well as rewarding
pursuits of high energy nuclear physics today. The observations of 
jet-quenching~\cite{jet1,jet2} and
the elliptic flow~\cite{fl1,fl2}, and  the success of the  partonic
recombination~\cite{rec} as a model for
hadronization in recent
experiments at the Relativistic Heavy Ion Collider at Brookhaven presage the
shift of the focus to even more interesting questions about how the plasma
 is formed
and how does it evolve. 

It is expected that the quantum statistical interference between
 identical particles emitted
from the relativistic heavy ion collisions will provide 
valuable inputs for these investigations.
 The pion intensity interferometry experiments
have however thrown up some startling questions, which need to be answered
before we can address these issues. The hydrodynamics models
which have traditionally 
provided a quantitative description to data on particle spectra 
and elliptic flow, fail
to explain the near identity of the so-called outward and side-ward
radii seen in these experiments and invariably lead to outward radii 
which are up to a factor of 2 larger than the side-ward radii~\cite{puzzle}.

 An investigation of the structure and dimensions of the source in 
terms of photon interferometry should provide a more direct and valuable
 insight into these questions.
The advantages of using photons for such studies is well known~\cite{bs}; they 
interact only weakly with the system after their production and are free
from such distorting effects as the  re-scattering~\cite{joe} and 
Coulomb interactions, which have dogged these investigations
using identical hadrons. Photons are also emitted at every stage
of the collision dynamics and from every point in the system- and not only from
the freeze-out surface.  We note though that some studies
have postulated that pions
may be emitted not only from the surface  as normally 
assumed; they also escape the system continuously from all 
points in the system~\cite{grassi} with some escape probability. A large 
probability for this will however  be hard to justify, as the pions do 
interact strongly
with the hadroinc medium. Still this possibility along with the modifications
introduced due to final state interactions for hadrons, necessitates 
a great deal of further study before the radii
obtained from the intensity interferometry of hadrons are used to get the
spatial and temporal extent of the source in these collisions.

Theoretical investigations of the nature of the photon correlation function for
relativistic heavy collisions, where a quark-gluon plasma may be formed have 
been carried out by several authors~\cite{dks,ors}. An experimental
measurement is a far more difficult proposition though, 
due to the huge back-ground of
decay photons and a considerably smaller production of direct photons. 
Thus, so far only one experiment at cyclotron energies~\cite{yves} (
where direct photons mostly originate from bremsstrahlung of
 protons against neutrons), and
 a measurement by the WA98 collaboration~\cite{wa98_2} (where photons originate 
from hadronic reactions and possibly quark matter) 
for the central collision of lead nuclei at CERN SPS have been reported.

It is expected that such experiments  will have a higher possibility of 
success at RHIC and
LHC energies.  Firstly, the larger initial temperatures expected there will
lead to a larger production of direct photons. 
Secondly, a 
large suppression of high momentum pions due to the onset of 
jet-quenching will lead
to a reduction in the back-ground from the decay photons. Furthermore,
a large production of photons having high $k_T$ is expected from 
the pre-equilibrium
stage of the collision~\cite{bms_phot,fms_phot}, which can be reliably crafted 
using the parton cascade model~\cite{bms} and it has been shown to have a
very distinct distribution of the source~\cite{bms_hbt}. 

The early theoretical investigations~\cite{dks,ors} of photon interferometry
were also aimed at isolating photons coming from the quark matter from
those coming from the hadronic matter. These studies concentrated
on photons having high transverse momenta ($k_T \gg $ 1 GeV), which are
more likely to be emitted from the early hot and dense phase of the plasma.

In the present work we focus  our attention on the photons having low and
intermediate 
$k_T \approx $ 100 MeV -- 2 GeV. The photons at the lower end of the $k_T$
 range considered would have a dominant contribution
from the late stages of the hadronic phase.  This should reveal a source which 
is strongly affected by the  radial flow. 
The higher end of the $k_T$ should have 
 a leading contribution from the early hot and dense stage of the plasma, 
when the flow is still small and the temperature is large.

We also incorporate several improvements over the early
exploratory works~\cite{dks}. Firstly,
we use the complete leading order results for the production of photons from
the quark matter~\cite{guy}, which has important contributions from the
bremsstrahlung, and annihilation of off-shell quarks~\cite{patrick}.
Secondly, for hadronic reactions we use the state of the art results
from Turbide et al.~\cite{simon} along with inclusion of hadronic form-factors
at the vortexes, incorporating strange mesons. The dominant hadronic
bremsstrahlung process ($\pi \pi \rightarrow \pi \pi \gamma $) is
 included~\cite{dipali} for
low $k_T$ photons and double counting via $\pi \pi \rightarrow \rho \gamma $
and $\rho \rightarrow \pi \pi \gamma$ avoided by limiting the contributions
of the latter two processes to $E_\gamma >$  500 MeV. This should be adequate
till a more detailed and complete calculation is available. Finally,
we use a much richer equation of state for the hadronic matter, with the
inclusion of all the particles in the particle data book, having $M < $ 2.5 GeV.
Results of our model calculations are compared with the recent data
obtained by the WA98 experiment and predictions given for RHIC and LHC energies.

We shall see that the competition between the cooling due to expansion
 and the blue-shift of
the photon spectra due to transverse expansion give rise to a unique
$k_T$ dependence of the outward and side-ward correlation radii, which evolves
rapidly as the initial temperature of the system changes, as we go from SPS
to RHIC to LHC.

 We first discuss the tools necessary for the analysis of the intensity
interferometry and briefly discuss the initial state and the rate of 
production of photons. Next we check whether the description used by us
provides a reasonable description of the single photon spectra measured
at . The results for some typical momenta of the photons and transverse
momentum dependence of correlation radii are discussed next. Finally we
give our conclusions.

\section{Formulation}

\subsection{The correlation function}

The spin averaged intensity correlation between two photons with
 momenta $\bf{k_1}$ and $\bf{k_2}$, emitted from a completely chaotic source,
is given by:
\begin{equation}
C(\mathbf{q},\mathbf{K})=1+\frac{1}{2}
\frac{\left|\int \, d^4x \, S(x,\mathbf{K})
e^{ix \cdot q}\right|^2}
                        {\int\, d^4x \,S(x,\mathbf{k_1}) \,\, \int d^4x \,
S(x,\mathbf{k_2})}
\label{def}
\end{equation}
where $S(x,\bf{k})$ is the space-time emission function, and
\begin{equation}
\mathbf{q}=\mathbf{k_1}-\mathbf{k_2}, \,\, \,
 \mathbf{K}=(\mathbf{k_1}+\mathbf{k_2})/2\, \, .
\end{equation}

In  a hydrodynamics calculation, the space-time emission function $S$ is
replaced by the rate of production of photon $EdN/d^4x d^3k$,
in the hadronic or the quark matter, as discussed earlier.

The results for the correlation function $C(\bf{q},\bf{K})$ are best
discussed in terms of the
 so-called outward, side-ward,  and longitudinal
momentum differences. Thus for the four-momentum $k_i^\mu$ of the $\it{i}$th
photon, we have,
\begin{equation}
k_i^\mu=(k_{iT}\, \cosh \, y_i,\mathbf{k_i})
\end{equation}
with
\begin{equation}
\mathbf{k_i}=(k_{iT}\,\cos \psi_i,k_{iT} \sin \psi_i,k_{iT} \, \sinh\, y_i)
\end{equation}
where $k_T$ is the transverse momentum, $y$ is the rapidity, and $\psi$ is the
azimuthal angle. Defining the difference and the average of the transverse
momenta,
\begin{equation}
\mathbf{q_T}=\mathbf{k_{1T}}-\mathbf{k_{2T}} \, \, ,
\mathbf{K_T}=(\mathbf{k_{1T}}+\mathbf{k_{2T}})/2\, ,
\end{equation}
we can write~\cite{sinyukov},
\begin{eqnarray}
q_{\text{long}}&=&k_{1z}-k_{2z}\nonumber\\
        &=&k_{1T} \sinh y_1 - k_{2T} \sinh y_2\\
\label{q_l}
q_{\text{out}}&=&\frac{\mathbf{q_T}\cdot \mathbf{K_T}}{K_T}\nonumber\\
       &=& \frac{(k_{1T}^2-k_{2T}^2)}
         {\sqrt{k_{1T}^2+k_{2T}^2+2 k_{1T} k_{2T} \cos (\psi_1-\psi_2)}}\\
\label{q_o}
q_{\text{side}}&=&\left|\mathbf{q_T}-q_{\text{out}}
       \frac{\mathbf{K_T}}{K_T}\right|\nonumber\\
        &=&\frac{2k_{1T}k_{2T}\sqrt{1-\cos^2(\psi_1-\psi_2)}}
      {\sqrt{k_{1T}^2+k_{2T}^2+2 k_{1T} k_{2T} \cos (\psi_1-\psi_2)}} \, .
\label{q_s}
\end{eqnarray}
The corresponding radii are obtained by approximating:
\begin{eqnarray}
 C(q_{\text{out}}, q_{\text{side}},q_{\text{long}})=
1+\frac{1}{2} 
  \exp \left[-\left(q_{\text{out}}^2R_{\text{out}}^2+
q_{\text{side}}^2R_{\text{side}}^2
+q_{\text{long}}^2R_{\text{long}}^2\right)/2\right]
\label{actual}
\end{eqnarray}

Note that for this choice of parametrization for the
 correlation function $R_i^2=1/<q_i^2>$, where
$i={\text{out}}$, ${\text{side}}$ and ${\text{long}}$, and the average
is performed over the distribution $(C-1)$.

A one dimensional analysis of the correlation function $C$ is some-times
performed in terms of the invariant momentum difference 
\begin{equation}
C(\sqrt{q_{\text{inv}}^2})=1+\frac{1}{2}
\exp\left[-q_{\text{inv}}^2 R_{\text{inv}}^2/2\right].
\label{one_1}
\end{equation}
where
\begin{eqnarray}
q_{\text{inv}}&=& \sqrt{-(k_1^\mu-k_2^\mu)^2}
 = \sqrt{-q_0^2+\mathbf{q}^2}\nonumber\\
&=&\sqrt{2 k_{1T}k_{2T} \left[ \cosh (y_1-y_2)-\cos (\psi_1-\psi_2) \right]}.
\end{eqnarray}

 However the significance of the corresponding
radius does not have a clear meaning~\cite{baker}, and it is also
not useful for comparing results obtained using different particles.
The variable
$\mathbf{q}^2 +q_0^2=q_{\text{inv}}^2+2q_0^2$ is also occasionally used for
 a one dimensional analysis~\cite{beker}, and one writes:
\begin{equation}
C(\sqrt{\mathbf{q}^2+q_0^2})=1+\frac{1}{2}
\exp\left[-(\mathbf{q}^2+q_0^2)R^2/2\right].
\label{one_2}
\end{equation}
Note also
 that $\mathbf{q}^2=q_{\text{out}}^2+q_{\text{side}}^2+q_{\text{long}}^2$.

Here, it is useful to realize that in actual studies of photon interferometry
the correlation function $C$ would be parametrized as:
\begin{eqnarray}
 C(q_{\text{out}}, q_{\text{side}},q_{\text{long}})= 1+\frac{1}{2} \times
\lambda \times
 \exp \left[-\left(q_{\text{out}}^2R_{\text{out}}^2+
q_{\text{side}}^2R_{\text{side}}^2
+q_{\text{long}}^2R_{\text{long}}^2\right)/2\right]
\end{eqnarray}
where $\lambda$ is a measure of single (S) vs. decay photons (D), under the
assumption that the source is completely chaotic. 
Obviously the later
do not contribute to the correlation (the life-time of $\pi^0$ 
is $\sim 10^{-16}$ seconds) and thus:
\begin{equation}
C(q_{\text{out}}=0, q_{\text{side}}=0,q_{\text{long}}=0)=
 1+\frac{1}{2} \lambda
\end{equation}
so that 
\begin{equation}
\lambda= \frac{ S^2}{(S+D)^2} ~.
\end{equation}
and therefore using the measured values of the total photon yield ($S+D$) at the
momentum $\bf{K}$, we can get the yield of single photons from the 
intercept of $C$ on the $y$-axis (Peressounko~\cite{ors}).
 This can provide a check on the results
on single photon obtained by, say, a subtraction of photons from decay of 
pions and etas. 

\begin{figure}[tb]
\centerline{\epsfig{file=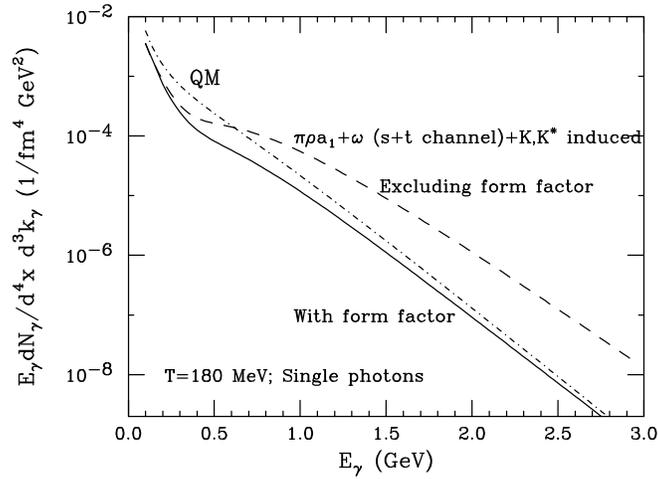,width=8.6cm}}
\caption{The  rate of emission of single photons from
hadronic reactions considered in Ref.~\cite{simon} at $T=$ 180 MeV,
with the inclusion (solid curve) of hadronic vortex form-factors.}
\label{fig1}
\end{figure}

\begin{figure}[tb]
\centerline{\epsfig{file=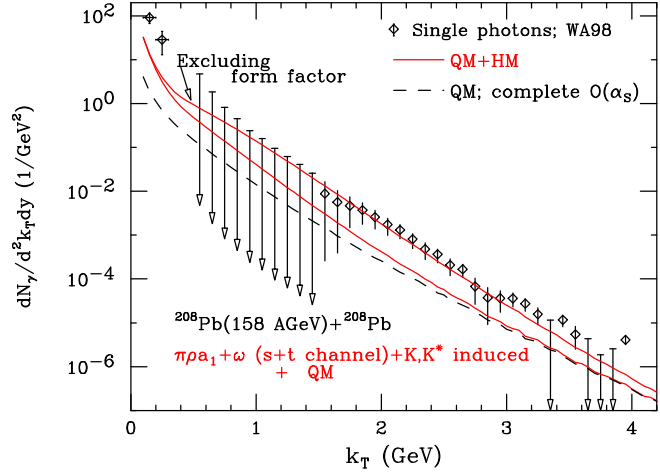,width=8.6cm}}
\caption{(Colour On-line)
The production of thermal photons from the quark matter plus 
hadronic matter, with and without the inclusion of the hadronic
form-factors in central collision of lead nuclei at CERN  energies.} 
\label{fig2}
\end{figure}

\begin{figure}[tb]
\centerline{\epsfig{file=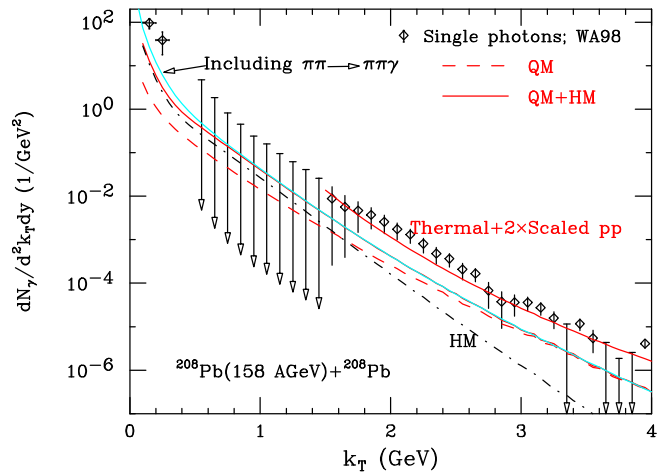,width=8.6cm}}
\caption{(Colour On-line)
The production of single photons in collision of lead nuclei
at CERN .}
\label{fig3}
\end{figure}

It is important to realize that this procedure is not directly 
applicable to the 
case where only a one-dimensional analysis in terms of $q_{\text{inv}}$ 
(Eq.\ref{one_1}) is performed.
The photon-pairs having transverse momenta in some bin around $K_T$, rapidity
difference within $\Delta y$, and azimuthal angle difference within
 $\Delta \psi$
 will lead to a range of outward, side-ward, and 
longitudinal momentum differences which will decide the actual 
correlation function $C$ (Eq.\ref{actual}).
Even a given value of $q_{\text{inv}}$ will admit a range of these 
momentum-differences, and thus the value of one-dimensional $C$ will be obtained
 by taking
an average over the corresponding results. This will lead to an 
effective $\lambda$
which will not reduce to unity, even if there were 
no decay-photons~\cite{dks_lamb}.
This ``effective'' $\lambda$, as well as the $R_{\text{inv}}$ will 
depend upon $K_T$, $\Delta K_T$,  $\Delta y$, $\Delta \psi$,
and the correlation radii, $R_{\text{out}}$, $R_{\text{side}}$, 
and $R_{\text{long}}$. Thus the direct photon yield can be measured only 
by assigning a value to these radii and becomes model dependent. Of-course, 
$R_{\text{inv}}$ can be assigned, though its meaning remains unclear and it
remains model dependent.

 A simple instance should exemplify this issue.
Consider a case where $y_1=y_2=0$ and $\psi_1=\psi_2$ so that 
$q_{\text{side}}=q_{\text{long}}\equiv$0 (see Eq.(\ref{q_l}-\ref{q_s})). 
In this case
$q_{\text{out}}=(k_{1T}-k_{2T})$, but $q_{\text{inv}}\equiv$ 0. 
Thus the ``true''
correlation function (Eq.\ref{actual}) will be decided by the value of the
outward momentum difference and the outward correlation radius, while
naively one would expect the one dimensional correlation function 
(Eq.\ref{one_1}), to  reduce to 1.5.
Thus even for a case when there are no decay photons, we would need a
correction factor to account for the range of values 
admitted by the ``true'' correlation function.

 The procedure may however still  work for the one-dimensional 
analysis in terms of
the variable $\sqrt{q_0^2+\bf{q}^2}$ (Eq.\ref{one_2}), as  this vanishes only
when all the three momentum-differences, $q_{\text{out}}$, $q_{\text{side}}$ and
$q_{\text{long}}$ also vanish simultaneously. The meaning of the corresponding
radius and its relation to the outward, side-ward and longitudinal
correlation radii, remains unclear though.

While we are discussing the merits of the one dimensional analysis in
 terms of the
invariant momentum difference, $q_{\text{inv}}$, we recall that for 
typical cases,
the correlation function $C$ differs substantially from unity for momentum 
differences $ \leq $ 0.2 GeV. For studies using single photons this limit
has to be less than $m_{\pi^0}$, to stay clear of the $\pi^0$ peak. If $k_{iT}$ 
are large, choosing $\Delta y$ and $\Delta \psi$ as small, we can get very
small values of the side-ward, outward, and longitudinal momentum
differences. However as $q_{\text{inv}}\propto k_{T}$, small values 
for $q_{\text{inv}}$ can be
obtained only if $k_{iT}$ are quite small. For larger value of $k_{iT}$,
the necessary $\Delta y$ and 
$\Delta \psi$ bins would be too small to admit meaningful statistics.
Thus, it is no wonder that both the photon interferometry experiments reported
in the literature so far~\cite{yves,wa98_2}, use the one-dimensional analysis 
in terms of $q_{\text{inv}}$ and utilize photons having very low $k_T$. We hope
that a much
larger statistics expected at RHIC and LHC will help us
get over this  problem
by providing results for the full 3-dimensional correlation function.

\subsection{Single photons}

 We very briefly recall the treatment for the space-time evolution of the
system and the production mechanism of the photons in these calculations.

For the first set of calculations,  we 
consider central collision of lead nuclei, corresponding to the
conditions realized at the CERN SPS. We further assume that a thermally and
chemically equilibrated quark-gluon plasma is produced in such collisions
at the initial time $\tau_0$, and use the assumption of is-entropic expansion
to estimate the initial temperature, $T_0$. Thus we have
\begin{equation}
\frac{2\pi^4}{45\zeta(3)} \frac{1}{A_T}\frac{dN}{dy}=4aT_0^3\tau_0 \, ,
\end{equation} 
where $A_T$ is the transverse area of the system, $dN/dy$ is the particle
rapidity density, and $a=42.25 \pi^2/90$ for a plasma of massless quarks
(u, d, and s) and gluons. The number of flavours for this purpose is
taken as $\approx$ 2.5 to account for the mass of the strange quarks. We
further assume a rapid thermalization of the plasma, constrained by the
uncertainty relation, $\tau_0=1/3T_0$. The initial energy density is
taken as proportional to the ``wounded - nucleon'' distribution, appropriate
for SPS energies.

 The quark-hadron phase transition is assumed to take
place at 180 MeV, and the freeze-out at 100 MeV. The relevant hydrodynamic
equations are solved under the assumption of boost-invariant longitudinal
and azimuthally symmetric transverse expansion using the procedure discussed
earlier~\cite{dks_hyd} and integration performed over the history of evolution. 
This procedure is known to give an accurate description of
 hadronic spectra~\cite{dks_had}.

We have already indicated the sources for evaluation of rates of photon
production from quark and hadronic matter.

As a first step we plot the rate of production of single photons due to all
the reactions among mesons considered in Ref.\cite{simon}. (see Fig.~\ref{fig1})
at $T=$ 180 MeV. We see that the inclusion of the form factor considerably
reduces the rate for production of photons having higher energies, leaving the 
results for the lowest energy photons unaltered. We also give the results
for the rates from quark matter at the same temperature. It should be remembered
that photon intensity interferometry should be less sensitive to the
specific details of the rates of production from quark and hadronic matter, 
but should
depend on their relative strengths.

In order to uniquely establish the importance of the mandatory requirement of
including the form-factors discussed in Ref.~\cite{simon}, we plot (see 
Fig.~\ref{fig2}) the sum of the production of thermal photons from the 
quark matter and the hadronic matter in a collision of lead nuclei at the
CERN SPS, with the initial conditions discussed earlier. The data
obtained by the WA98 experiments~\cite{wa98_2,wa98_1} and the production from
the quark-matter are also shown for a comparison. We immediately note
that the exclusion of the form-factors will considerably overestimate the
production of single photons at higher $k_T$, as it is well established from 
several studies that up to half of the single photons measured by the 
WA98 experiment have their origin in the prompt-QCD process, which should
be added to the thermal photons to get  a quantitative description
of the production of the single photons. We further note that the 
calculations underestimate the data by a factor of 7 -- 9 at the lowest
$k_T$ reported recently~\cite{wa98_2} (see discussion later).

Finally, in Fig.~\ref{fig3} we show our results for the production of
single photons, by summing the thermal contributions (with the inclusion of
form-factors while estimating the hadronic rates) and the prompt-QCD
contribution. For the later we use a parametrization of all the $pp$ 
data~\cite{dks_scale}
and scale it for collision of lead nuclei. We account for the intrinsic
$k_T$ of the partons by multiplying this scaled $pp$ contribution by a
factor of 2. We see that the sum of the thermal and prompt contributions
provides a satisfactory explanation of the single photon spectrum at large
$k_T$ in the present model.

We also note that while an inclusion of the pionic bremsstrahlung improves the
description of the data at the lower $k_T$ marginally, the theoretical
calculations are still well below the estimated (see later) experimental
results. We can think of two possible reasons for this shortfall.
One possibility is that there may be additional reactions, e.g., involving
baryons, which may contribute at low $k_T$. The other, and possibly more
plausible, reason could be the neglect of the pionic chemical 
potential~\cite{koch} in the present work, and which can arise
arise to-wards the end of hadronic phase~\cite{sean}.
%%%%%%%%%%%%%%%%%%%%%%%%%%%%%%% revised MS %%%%%%%%%%%%%%%%
Let us elaborate on this aspect. 
Recall that the pion-chemical potential ($\mu_\pi$)
may reach a value of about 60 MeV in collision of lead nuclei at SPS 
energies\cite{uli} when $T$ drops to 100 MeV. This will imply an enhancement
in the rate of production of photons from the hadronic matter by
approximately a factor of $\exp(\mu_\pi/T)$ for every pion in the entrance
channel, and also some increase due to Bose enhancement factor in the
exit channel, if there is a pion there.
%%%%%%%%%%%%%%%%%%%%%%%%%%%%%%%%%%%%%%%%%%%%%%%%%%%%%%%%%%%%%%%%
 These aspects
are under investigation. We may add however that the inclusion of these
corrections should not drastically alter the following results, as 
the hadronic phase contributions, already dominate the yield of
photons at low $k_T$, considerably.

\begin{figure}[tb]
\centerline{\epsfig{file=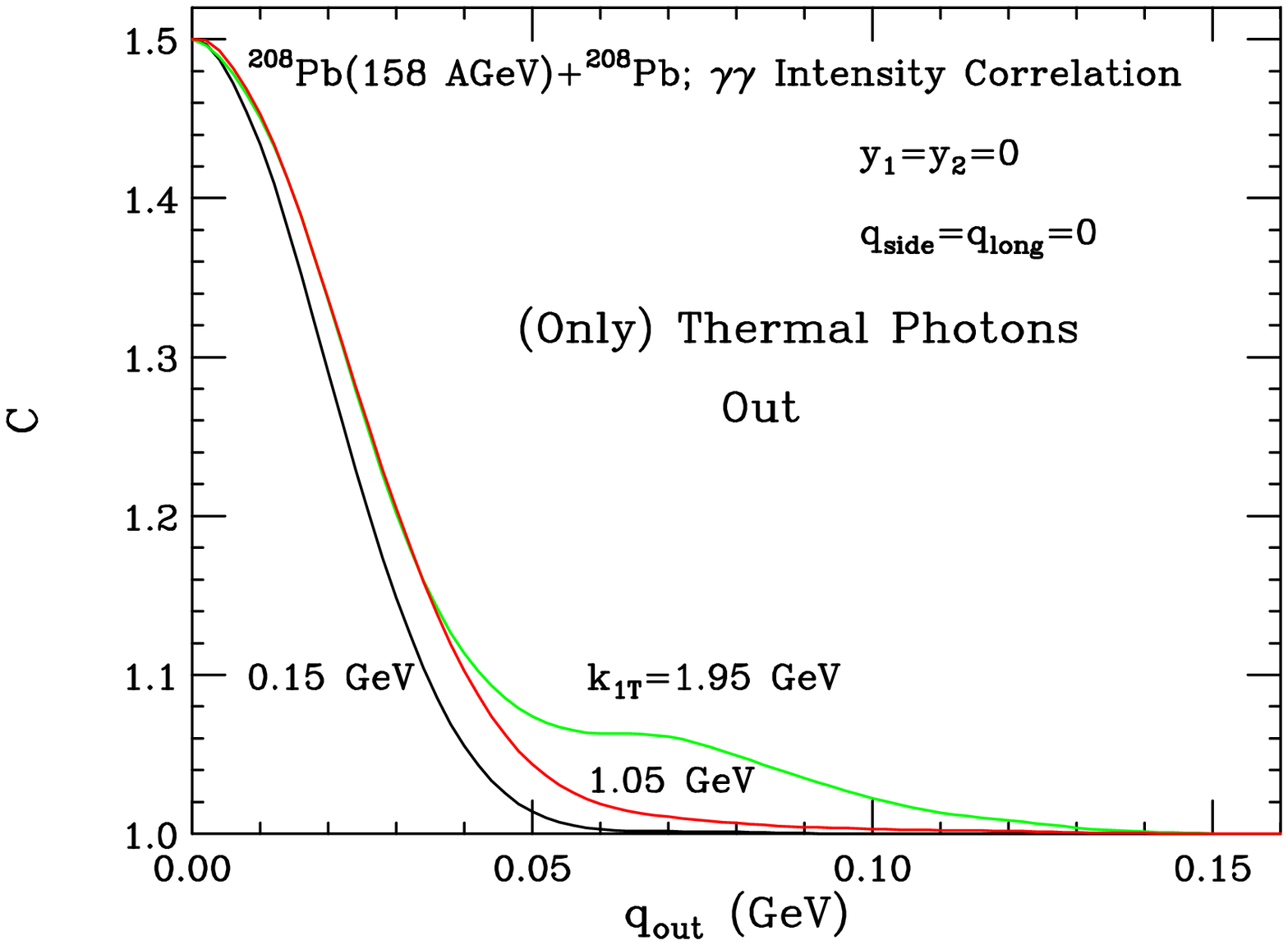,width=8.6cm}}
\centerline{\epsfig{file=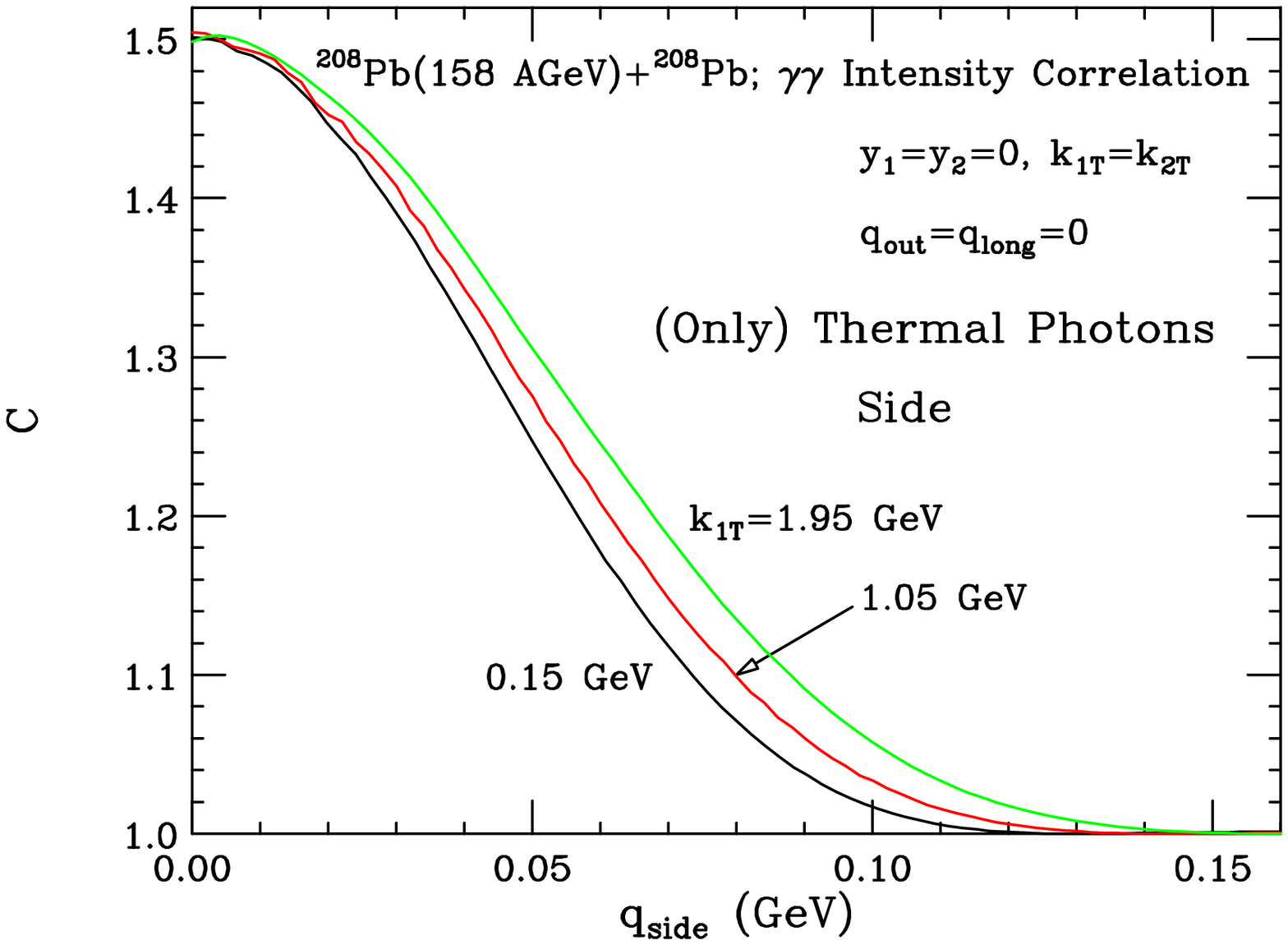,width=8.6cm}}
\centerline{\epsfig{file=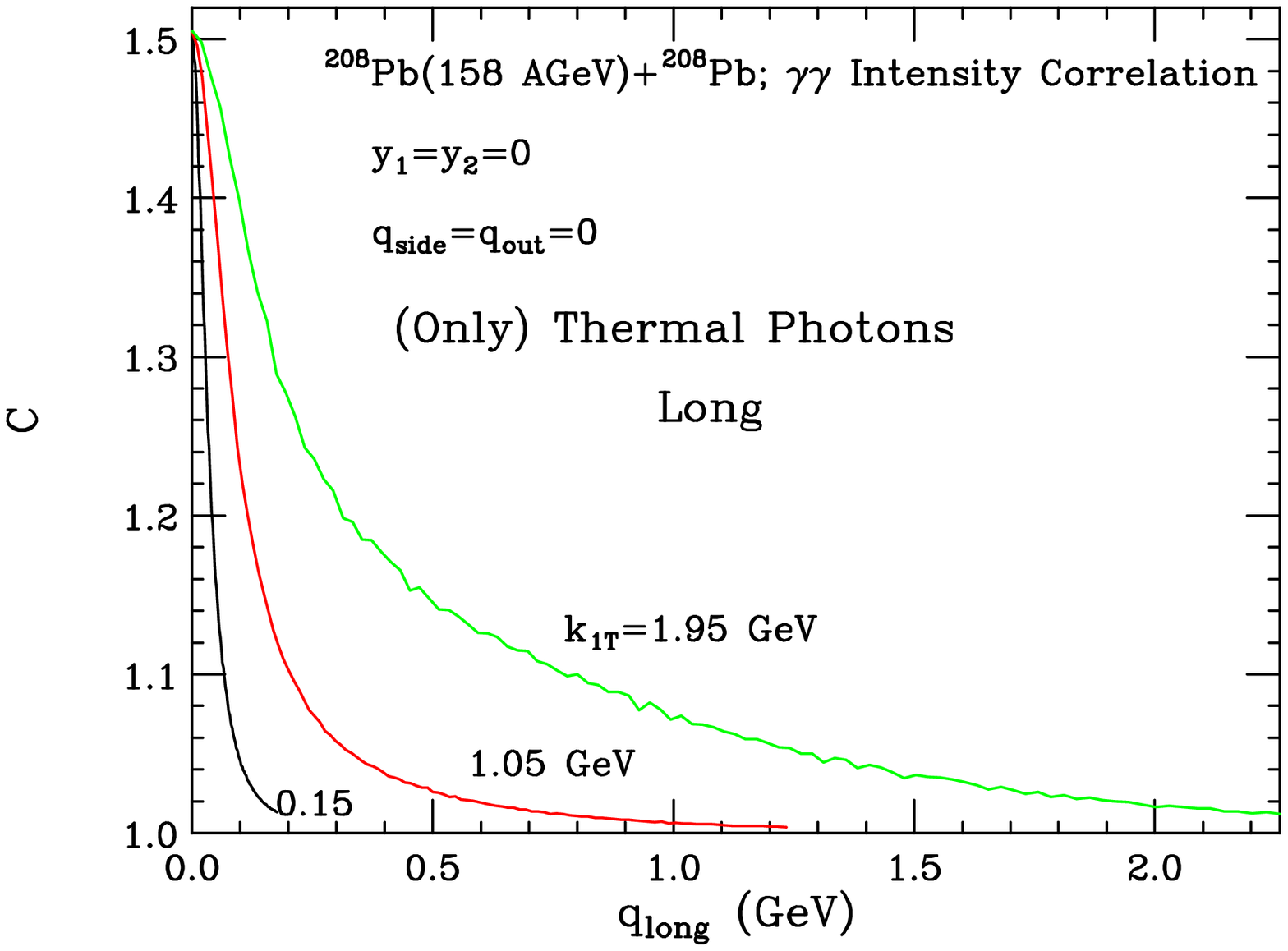,width=8.6cm}}
\caption{(Colour On-line)
The outward, side-ward, and longitudinal correlation function
for direct photons produced in central collision of lead nuclei at CERN .}
\label{fig4}
\end{figure}

\begin{figure}[tb]
\centerline{\epsfig{file=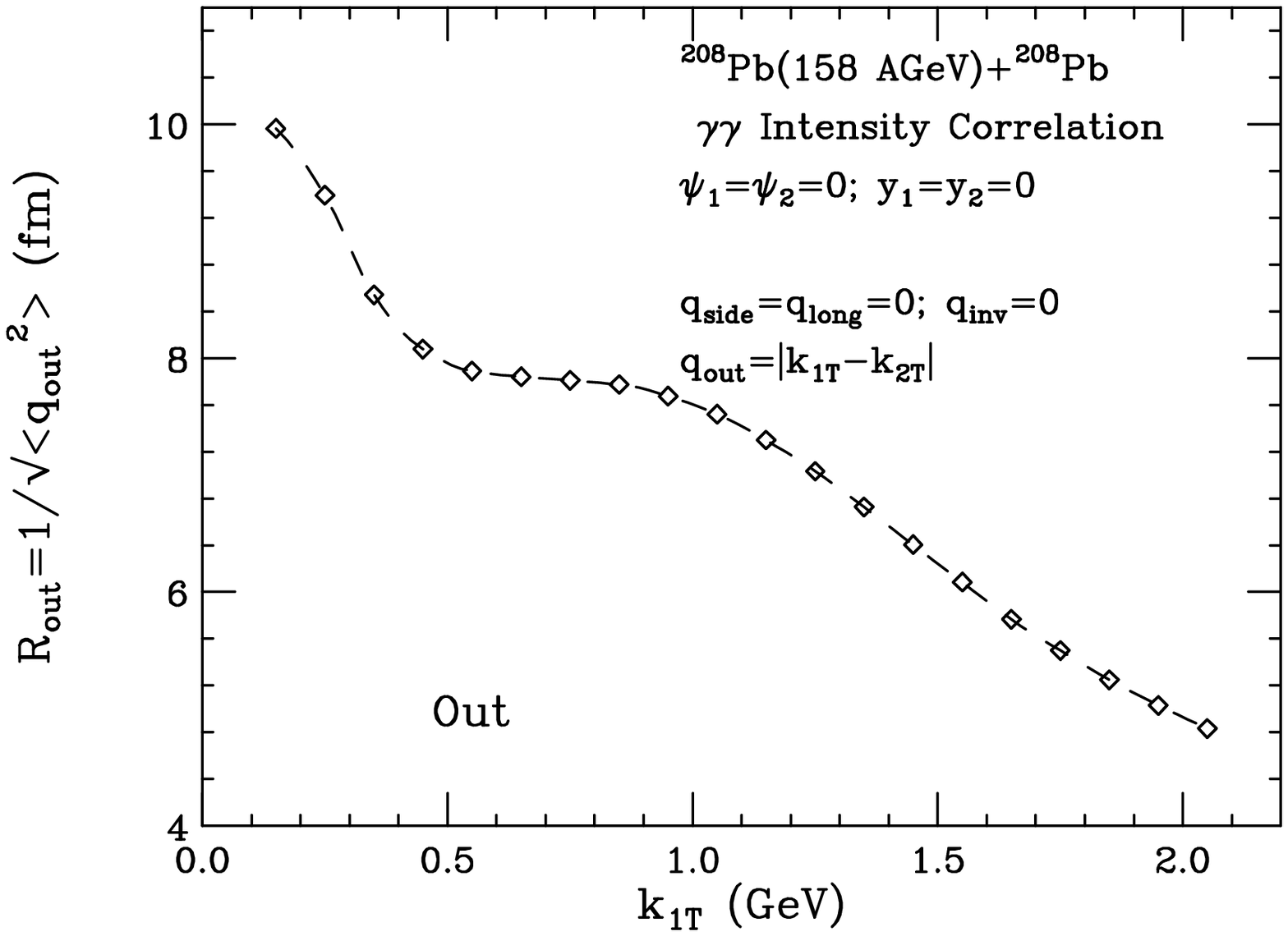,width=8.6cm}}
\centerline{\epsfig{file=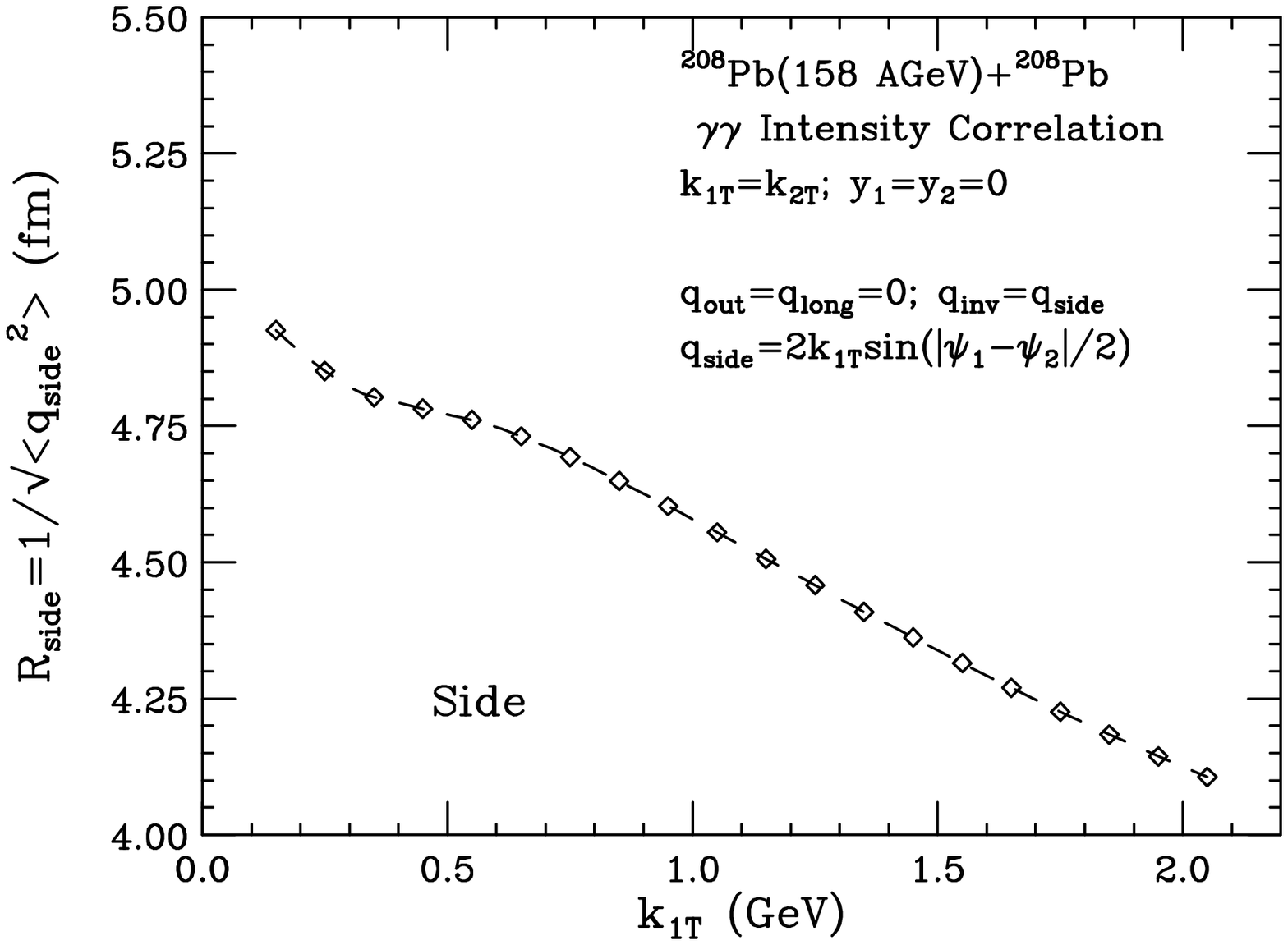,width=8.6cm}}
\centerline{\epsfig{file=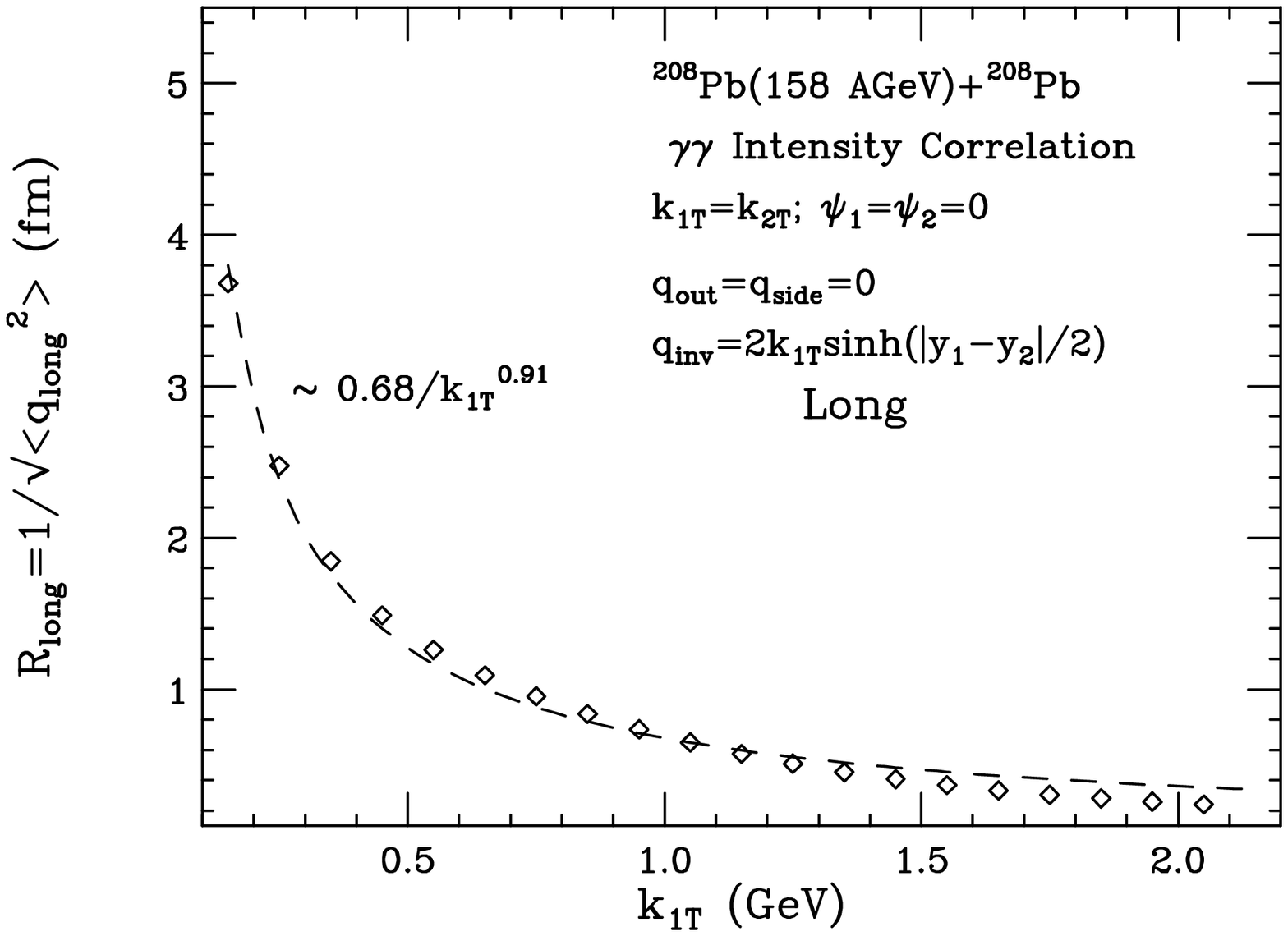,width=8.6cm}}
\caption{The transverse momentum dependence of outward, side-ward and 
longitudinal
radii for photons from central collision of lead nuclei at CERN .}
\label{fig5}
\end{figure}
\begin{figure}[tb]
\centerline{\epsfig{file=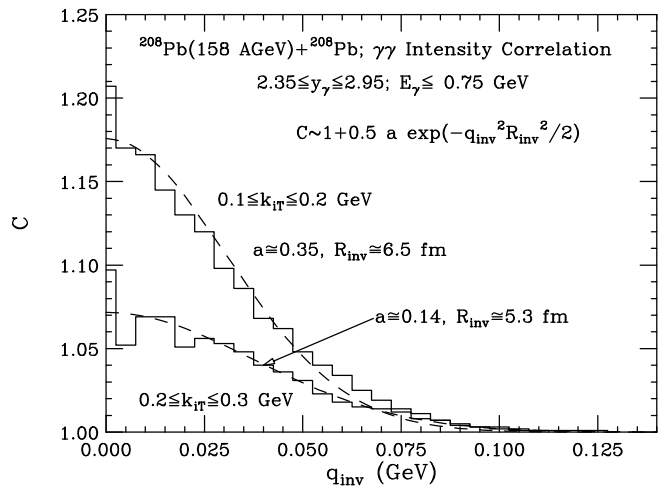,width=8.6cm}}
\caption{The one dimensional correlation function for the kinematic window
used in WA98 experiment~\cite{wa98_2}, assuming a fully source and emitting
only single photons. Central collision of lead nuclei is considered.}
\label{fig6}
\end{figure}

\begin{figure}[tb]
\centerline{\epsfig{file=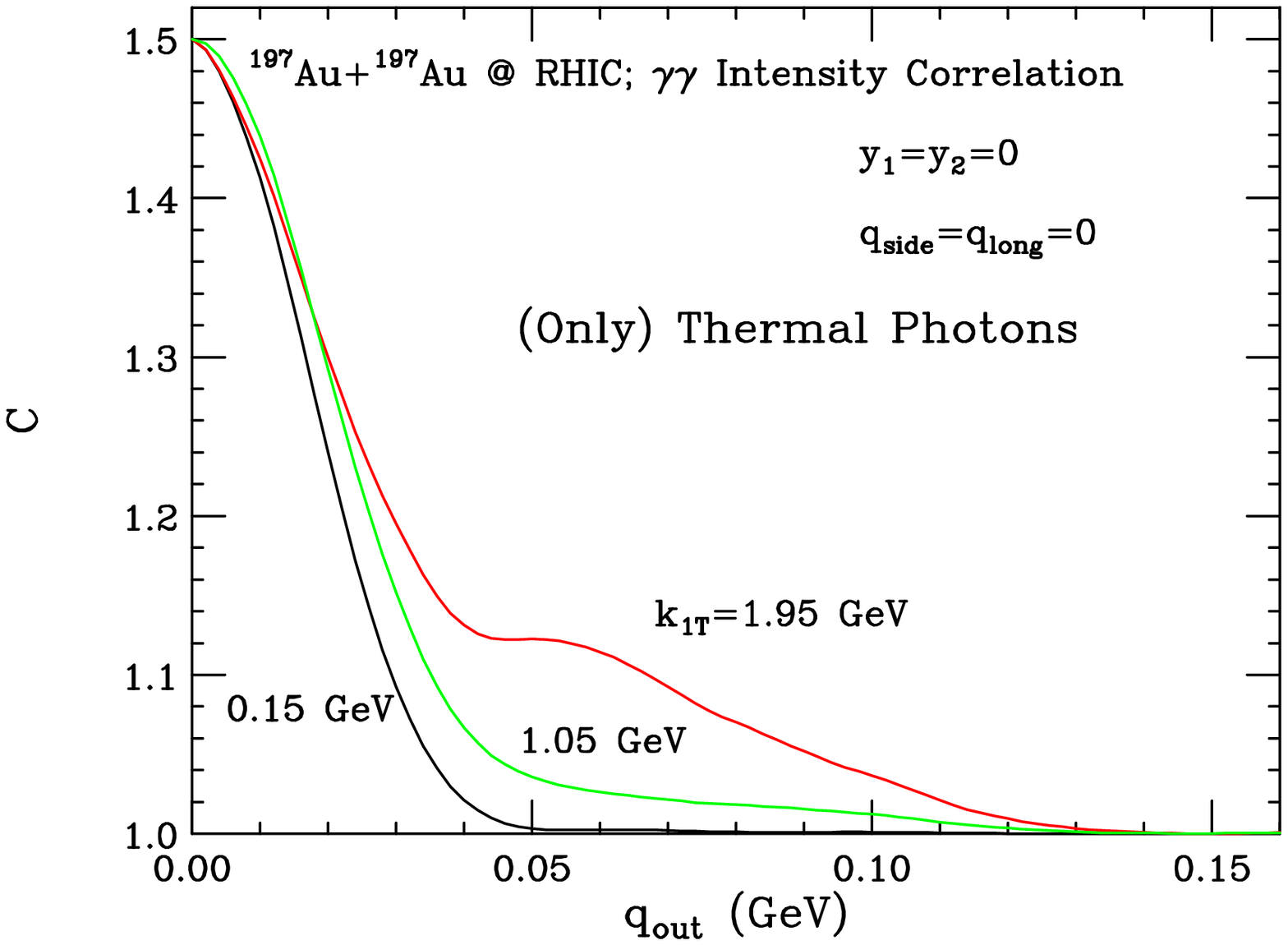,width=8.6cm}}
\centerline{\epsfig{file=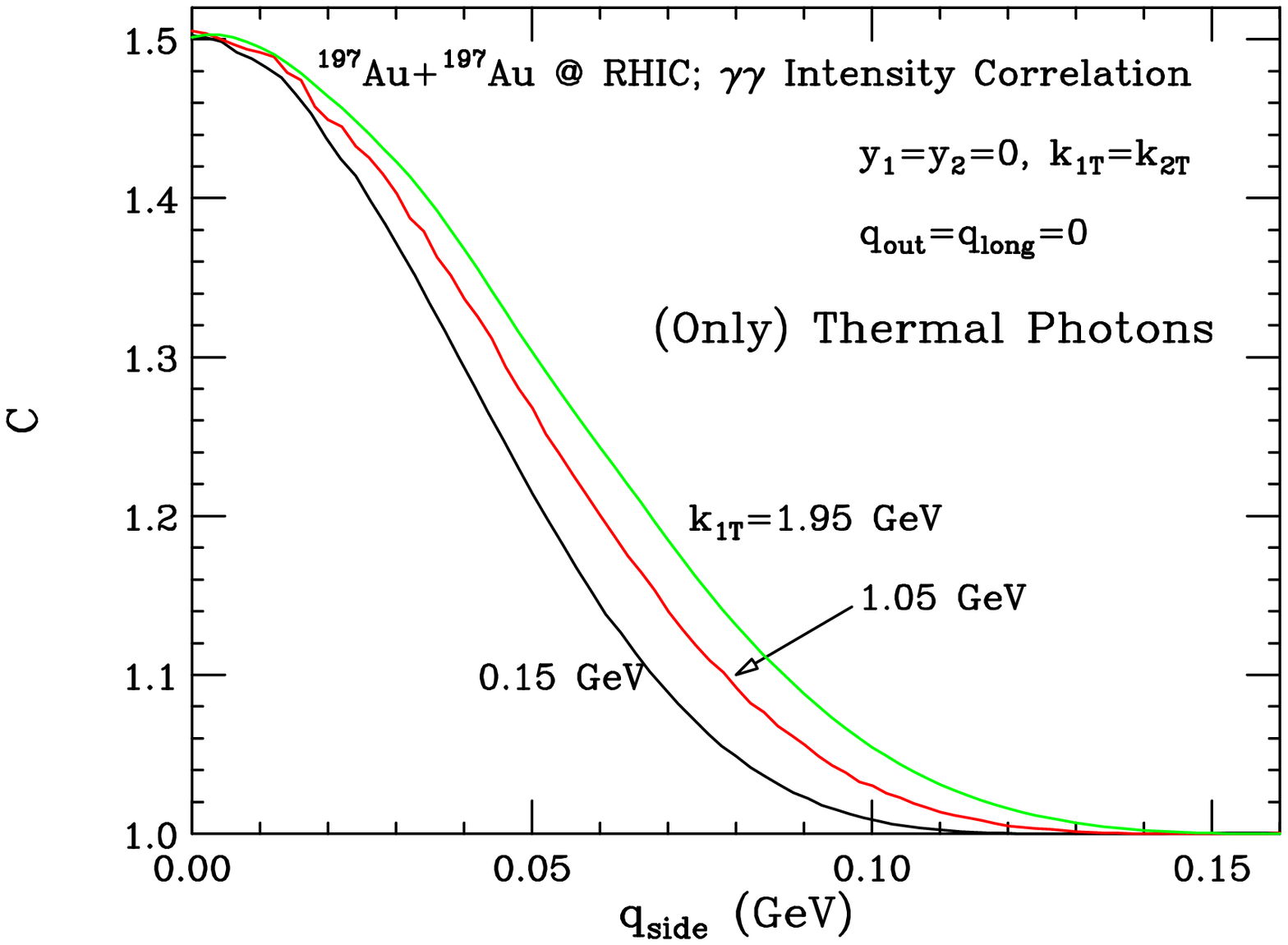,width=8.6cm}}
\centerline{\epsfig{file=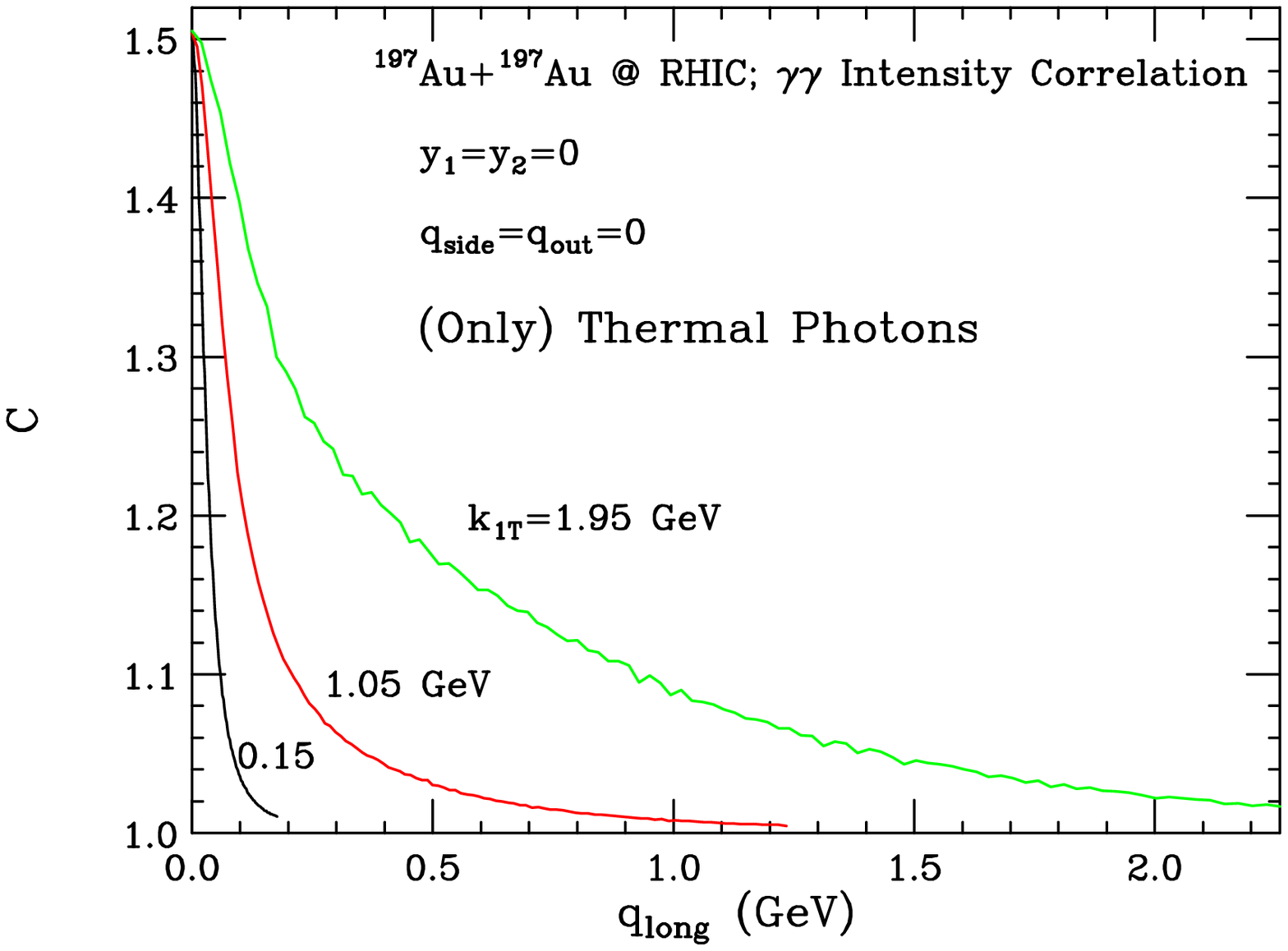,width=8.6cm}}
\caption{(Colour On-line)
The outward, side-ward, and longitudinal correlation function
for thermal photons produced in central collision of gold nuclei at BNL RHIC.}
\label{fig7}
\end{figure}

\begin{figure}[tb]
\centerline{\epsfig{file=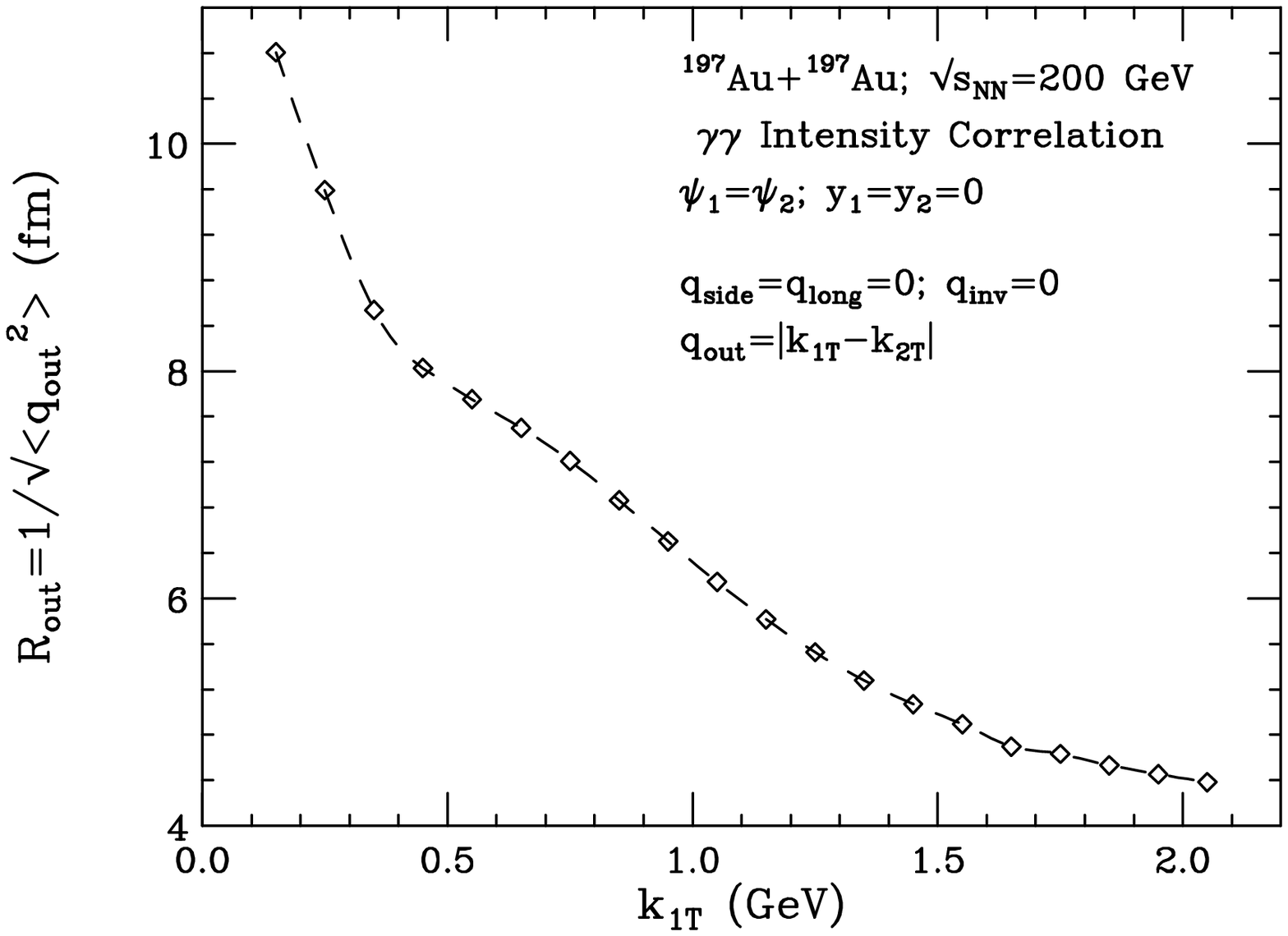,width=8.6cm}}
\centerline{\epsfig{file=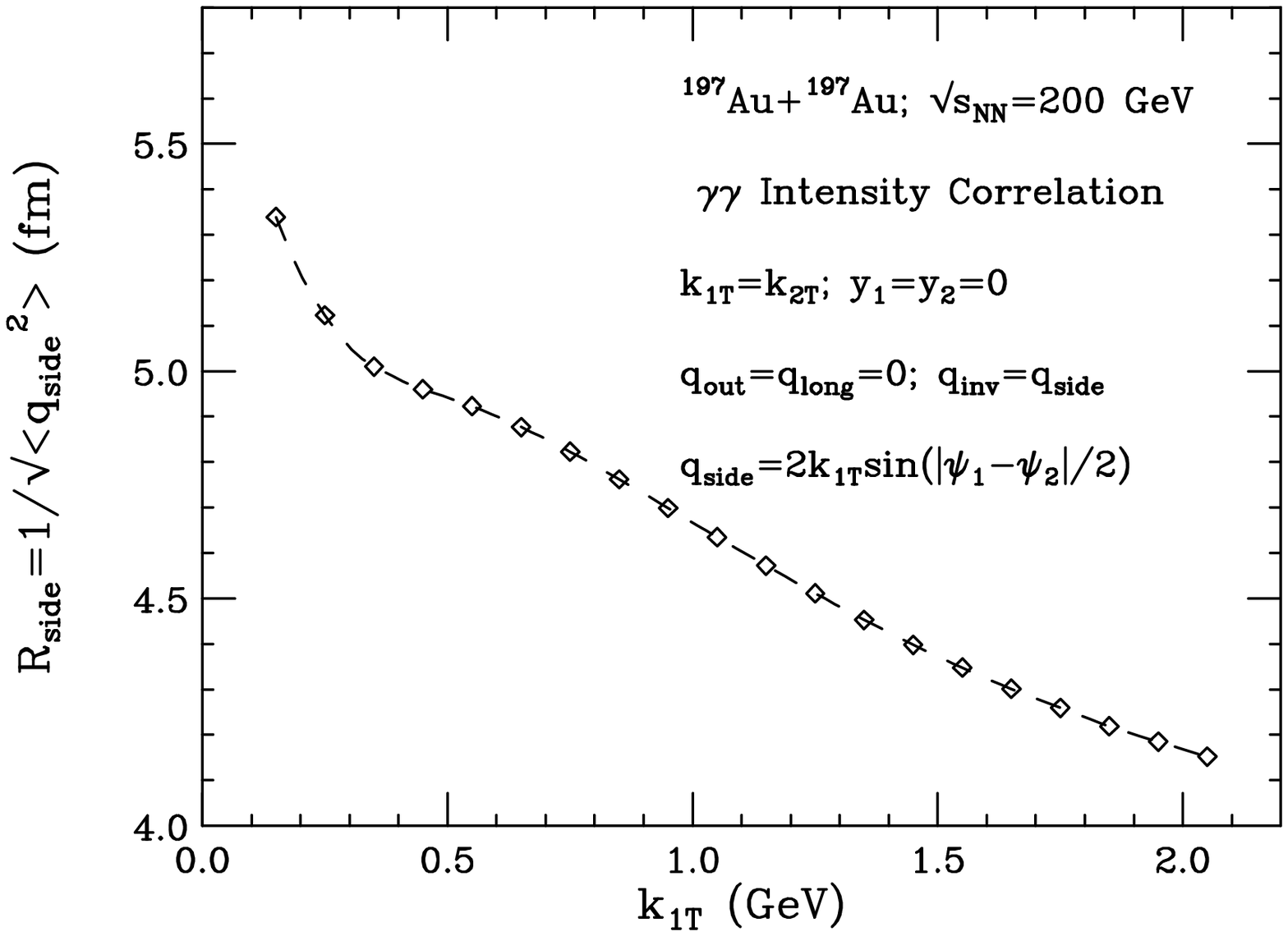,width=8.6cm}}
\centerline{\epsfig{file=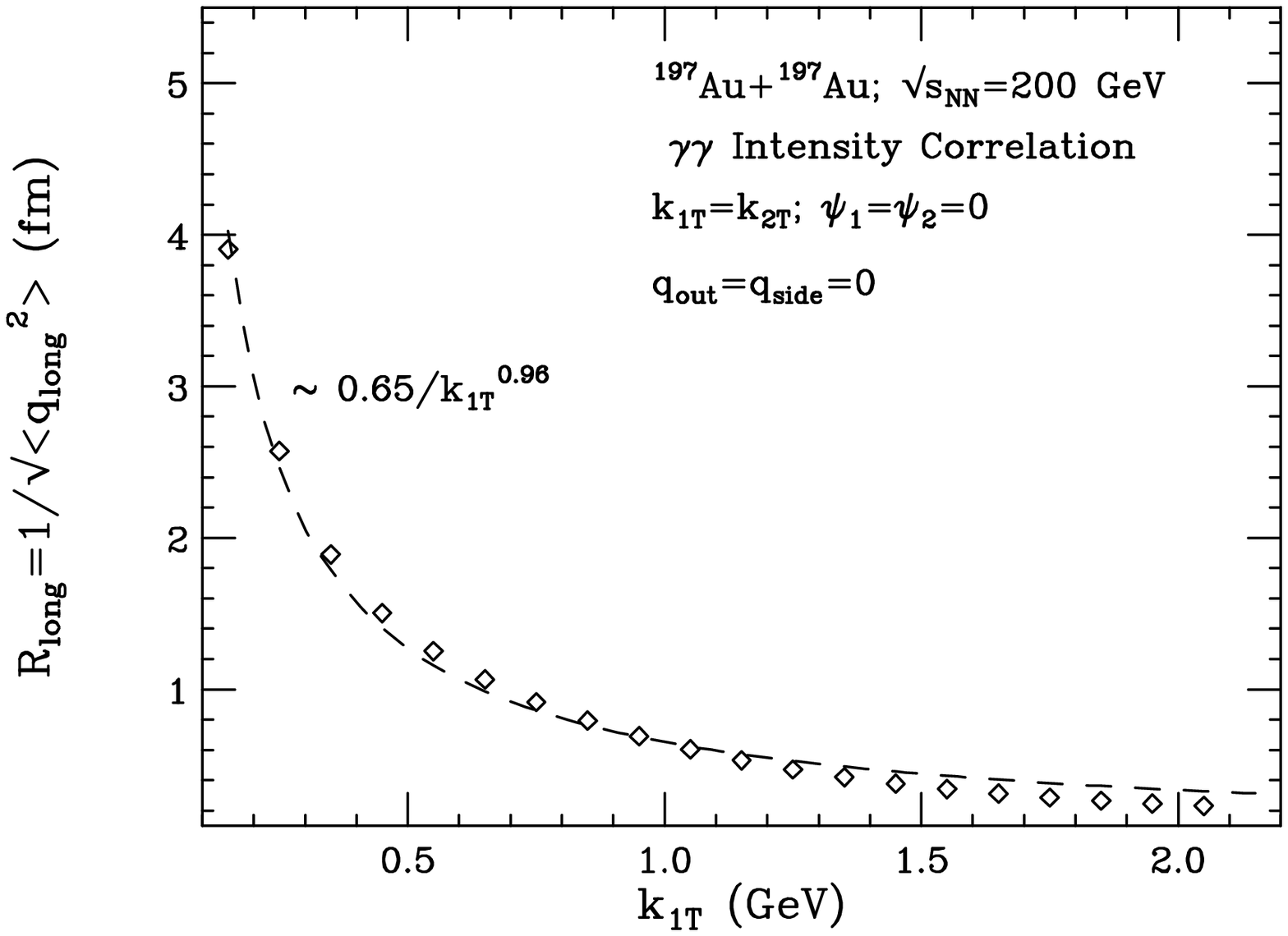,width=8.6cm}}
\caption{The transverse momentum dependence of outward, side-ward and 
longitudinal
radii for thermal from central collision of gold nuclei at BNL RHIC.}
\label{fig8}
\end{figure}

\begin{figure}[tb]
\centerline{\epsfig{file=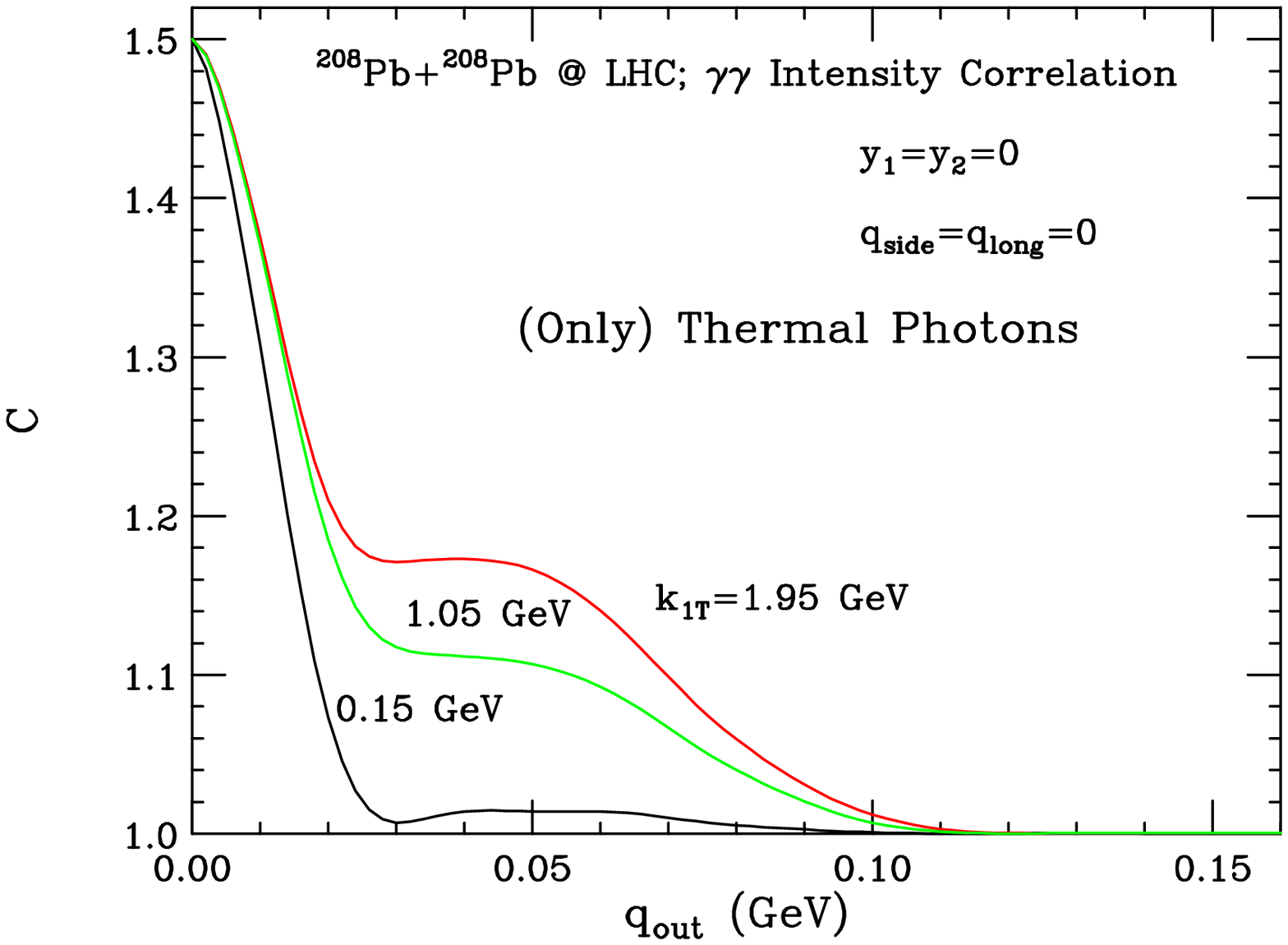,width=8.6cm}}
\centerline{\epsfig{file=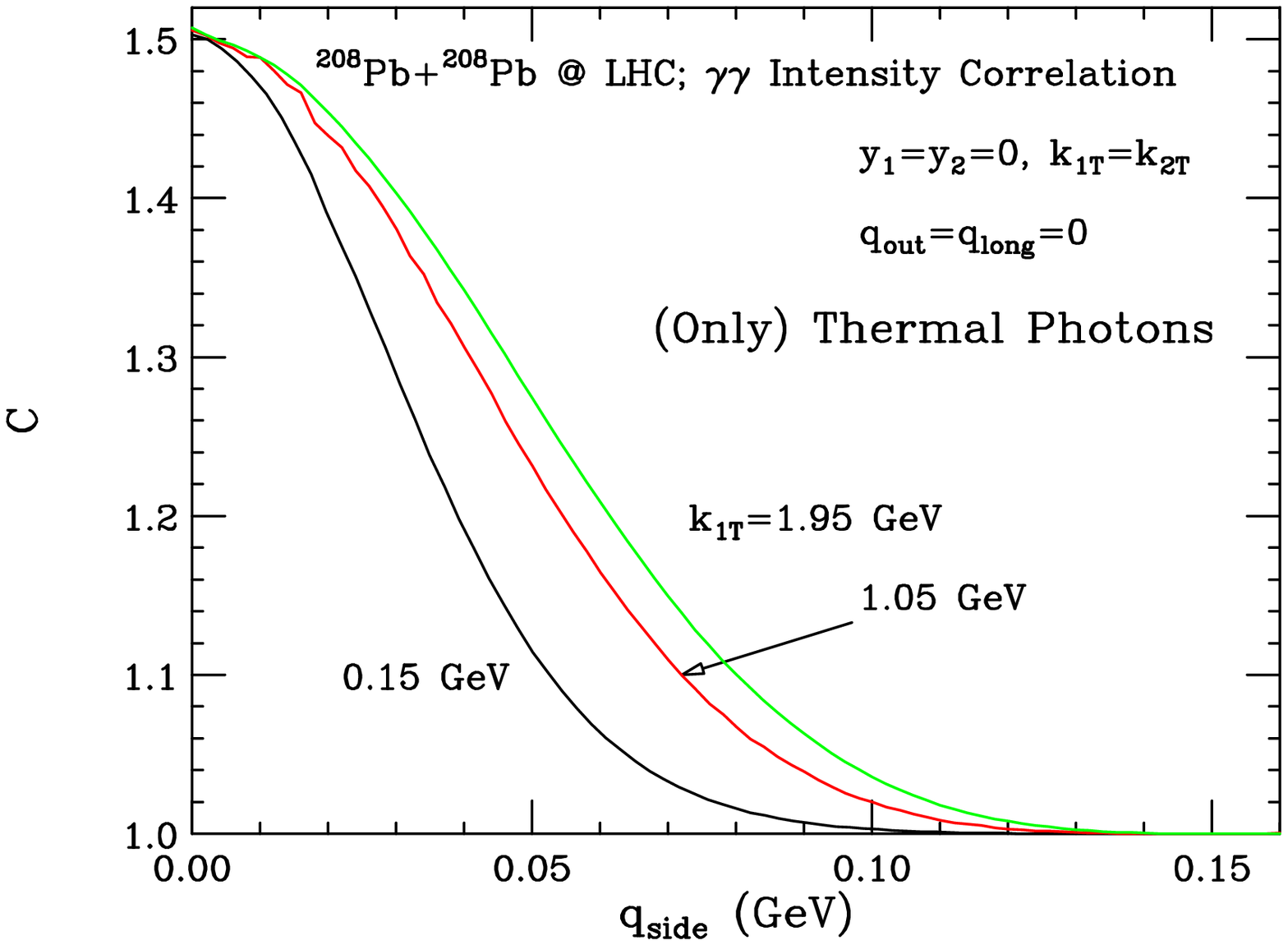,width=8.6cm}}
\centerline{\epsfig{file=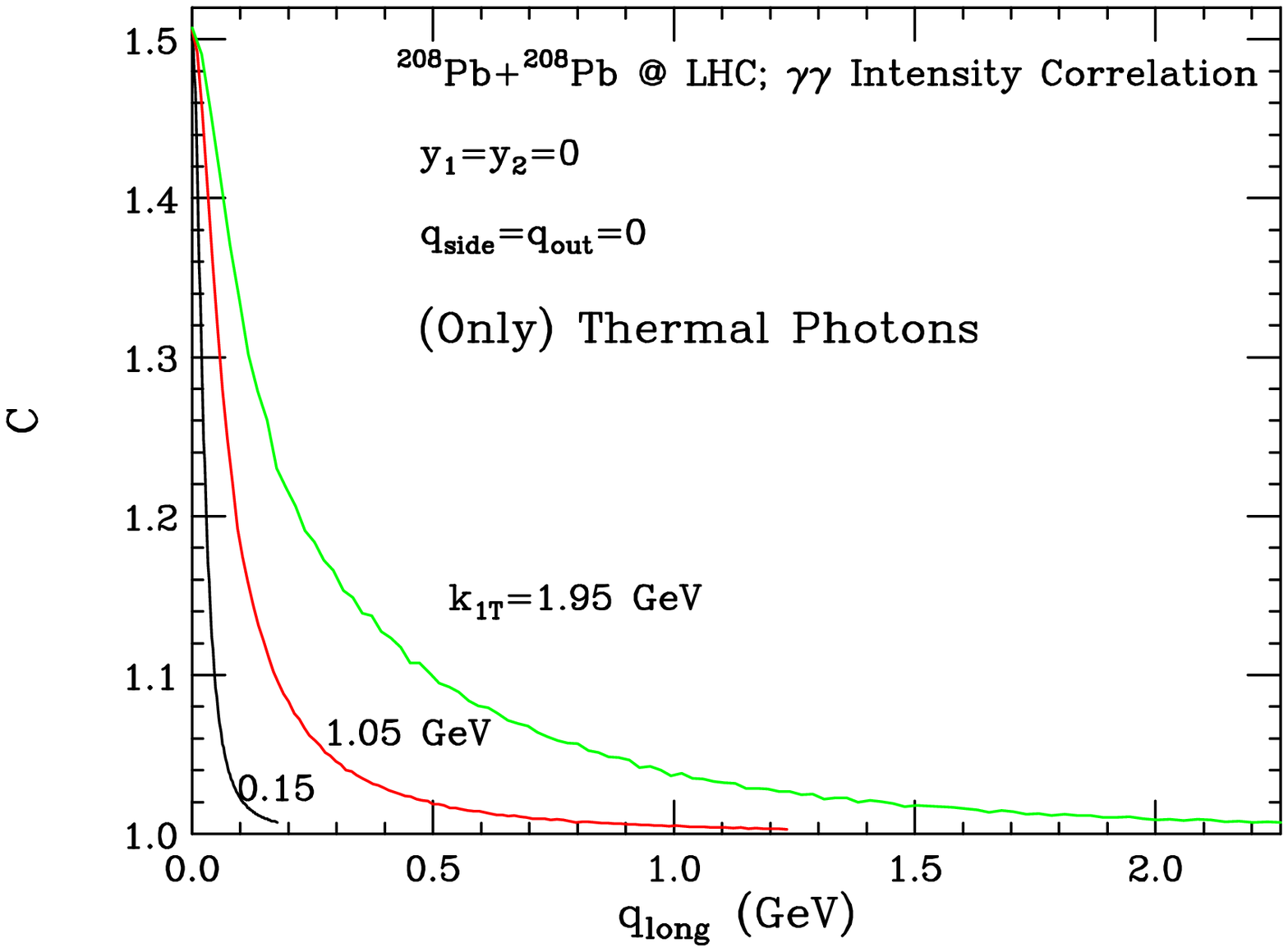,width=8.6cm}}
\caption{(Colour On-line)
The outward, side-ward, and longitudinal correlation function
for thermal photons produced in central collision of lead nuclei at CERN LHC.}
\label{fig9}
\end{figure}

\begin{figure}[tb]
\centerline{\epsfig{file=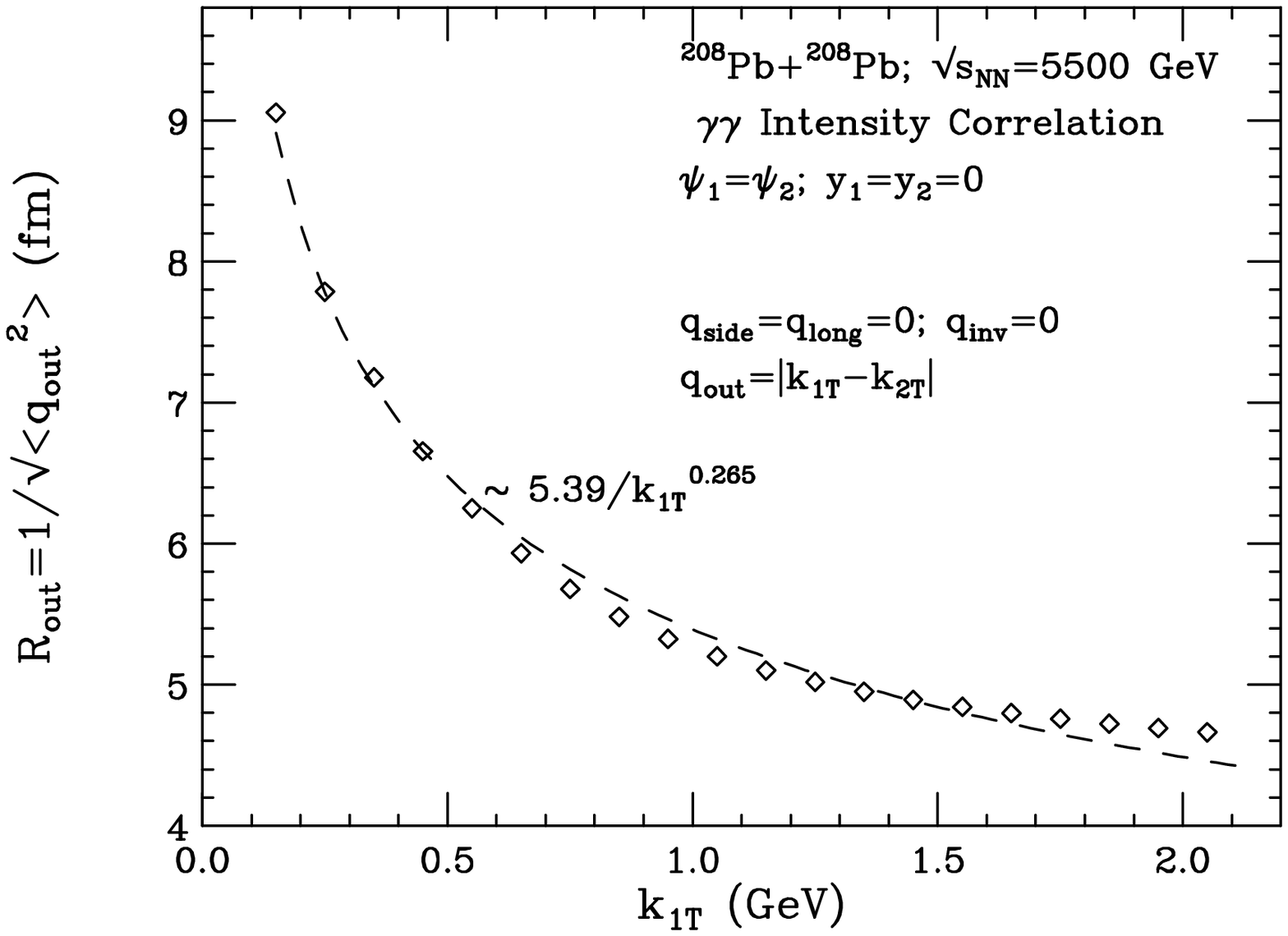,width=8.6cm}}
\centerline{\epsfig{file=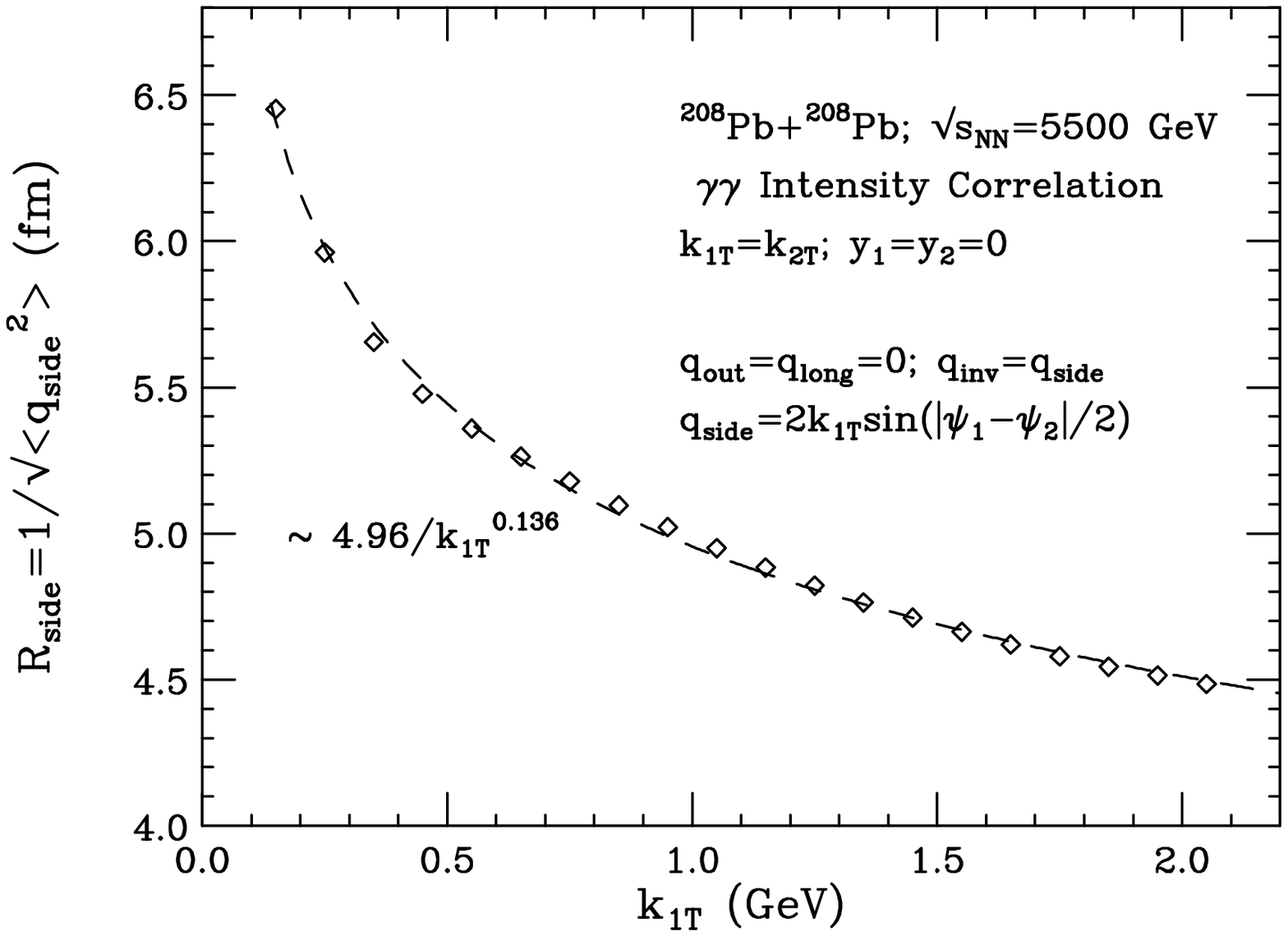,width=8.6cm}}
\centerline{\epsfig{file=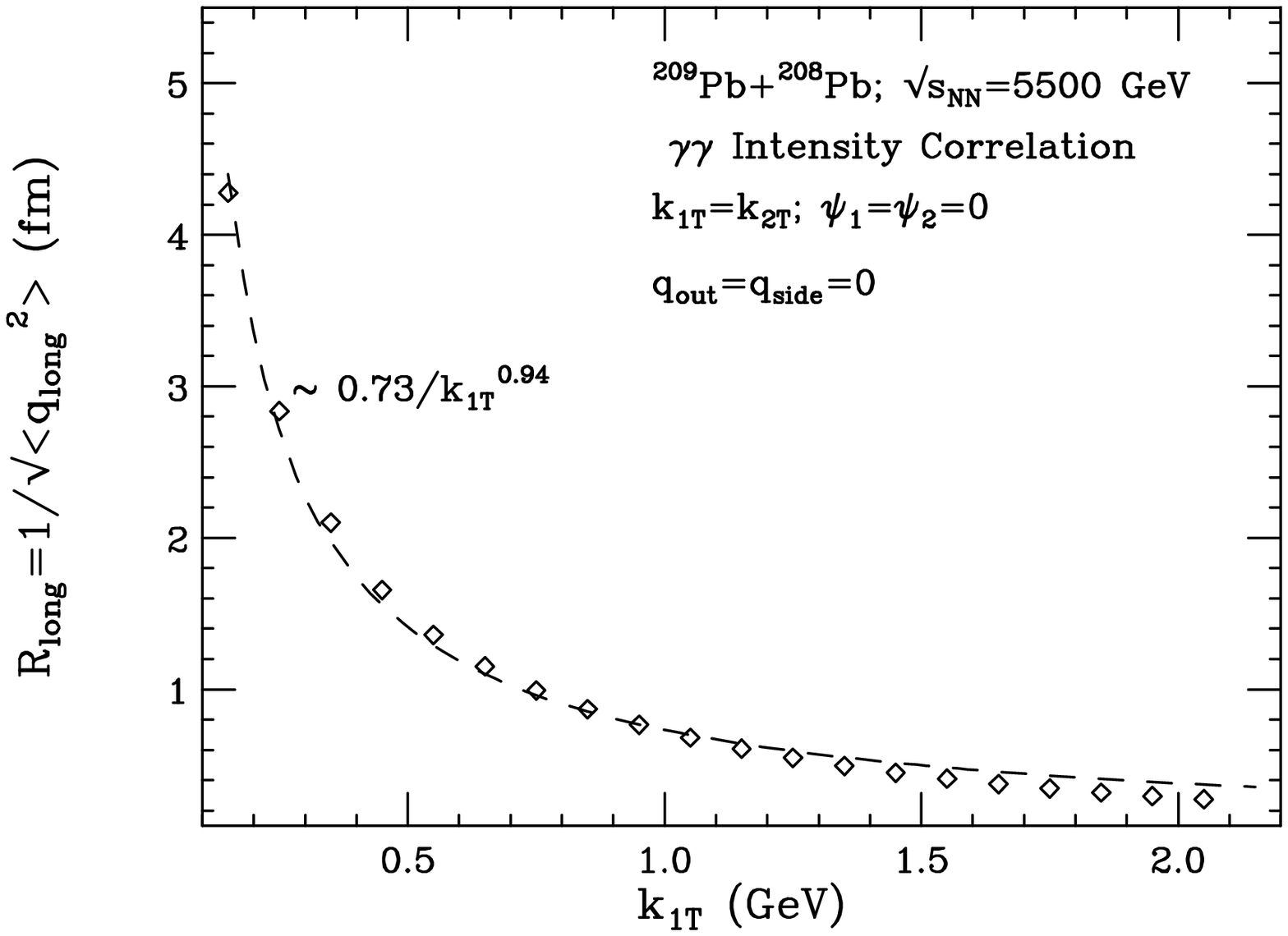,width=8.6cm}}
\caption{The transverse momentum dependence of outward, side-ward and 
longitudinal
radii for thermal from central collision of lead nuclei at CERN LHC.}
\label{fig10}
\end{figure}

The calculations at RHIC and LHC energies are performed in an analogous manner,
taking $dN/dy$ at $y=0$ as 1260 and 5625~\cite{LHC} respectively
 for central collisions of
gold (at RHIC) and  lead (at LHC) nuclei. This procedure has been discussed
repeatedly in literature; 
the only difference we have in the present work is that we use the
 state of the art results
for the rates, and a more complete equation of state for the hadronic
matter.

\begin{figure}[tb]
\centerline{\epsfig{file=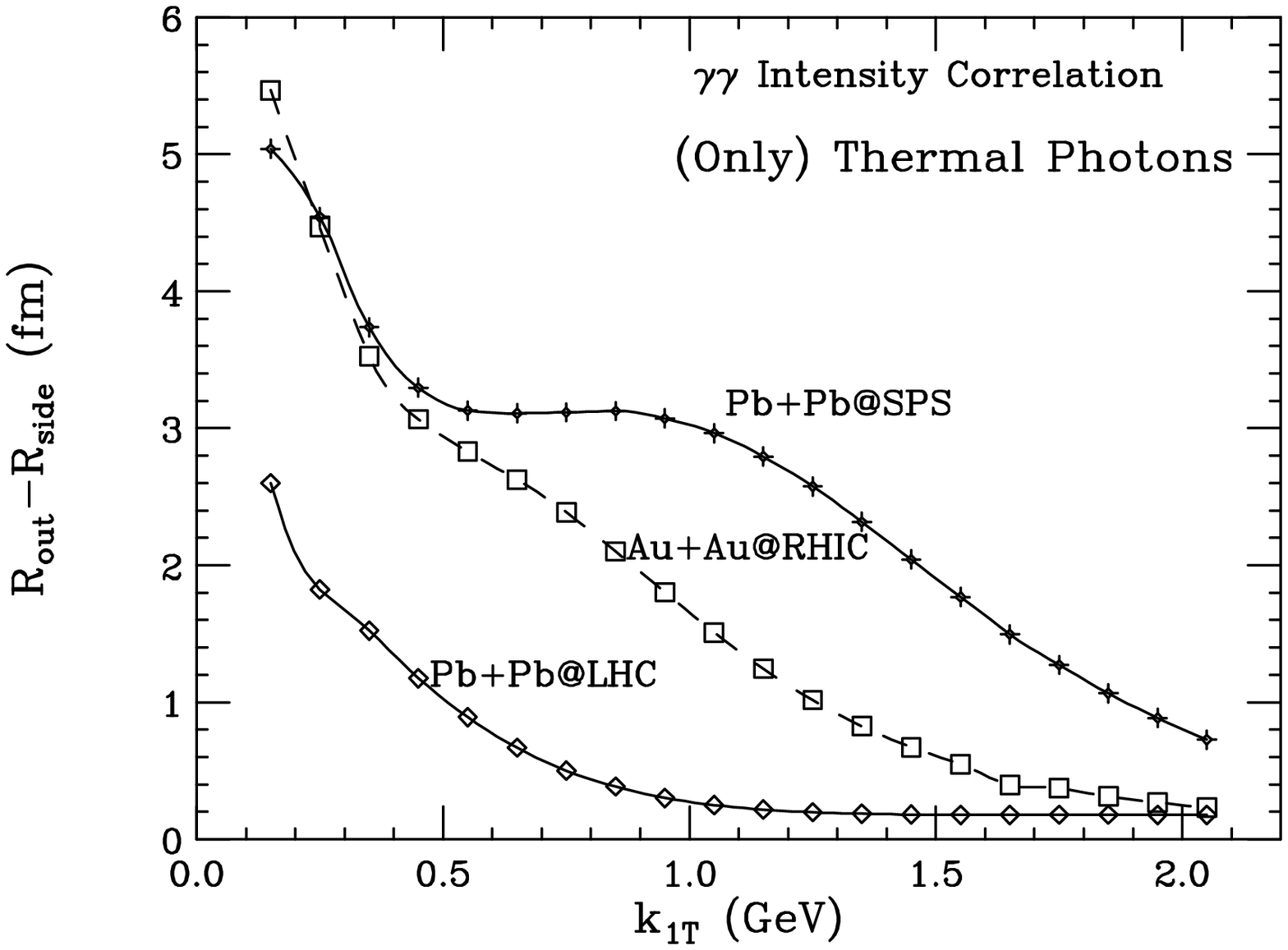,width=8.6cm}}
\centerline{\epsfig{file=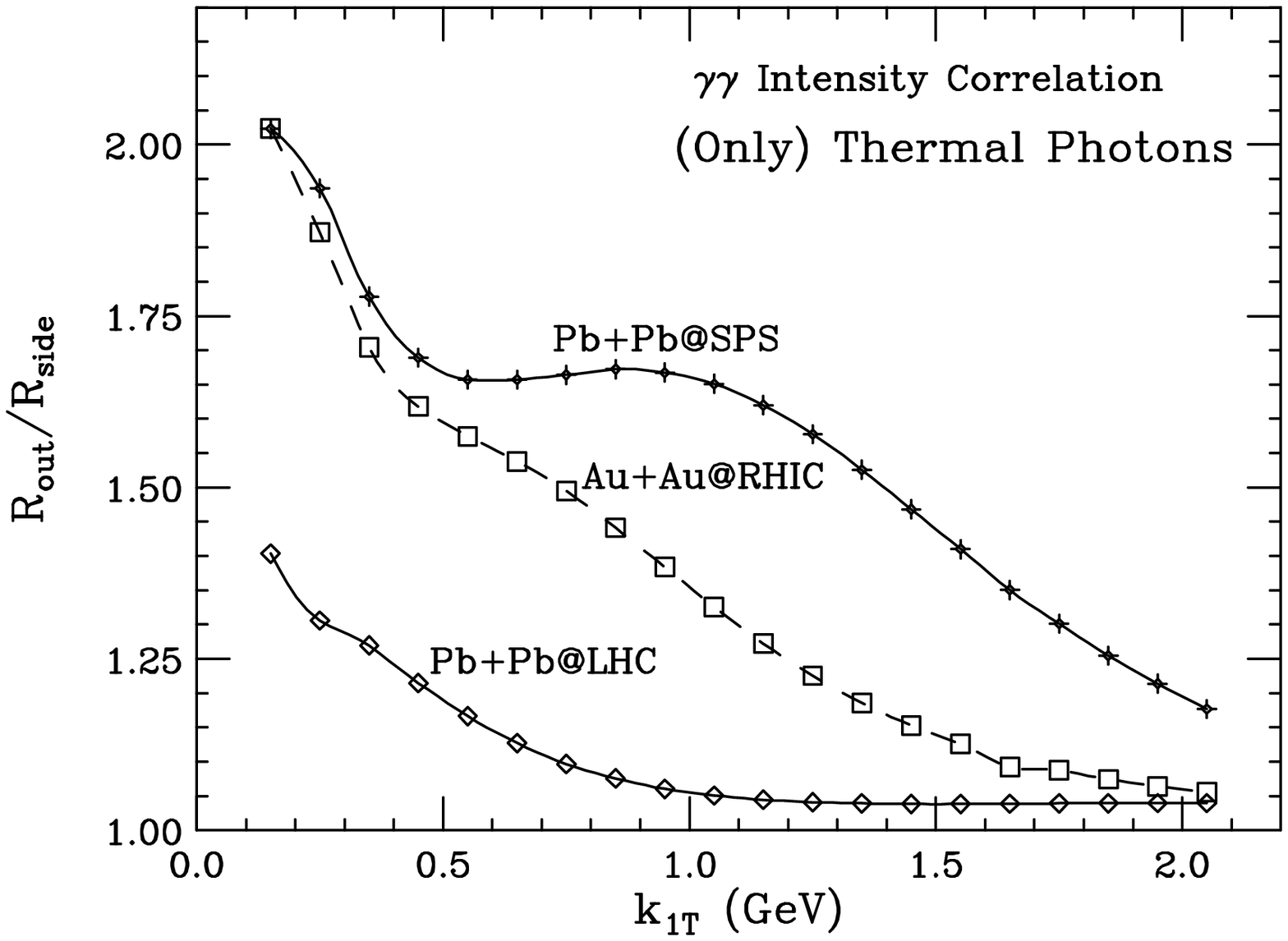,width=8.6cm}}
\caption{The difference of out-ward and side-ward correlation radii
for thermal photons at CERN SPS, BNL RHIC, and CERN LHC.}
\label{fig11}
\end{figure}

\subsection{Photon intensity interferometry at SPS energies}

As a first step, let us investigate the differences in the patterns of the
intensity interferometry of photons having high and low $k_T$. Thus we perform
calculations by choosing $k_{iT}$, $\psi_i$, and $y_i$ of photons such
that two of the three momentum differences, $q_{\text{out}}$, 
$q_{\text{side}}$, and $q_{\text{long}}$, vanish in turn. This theoretical
construction helps us obtain the corresponding correlation functions.

We show typical results for $k_{1T}=$ 0.15 GeV, 1.05 GeV, and 1.95 GeV
in Fig.~\ref{fig4}. We see a very interesting feature in the outward correlation
function: as the transverse momentum of the photon increases, we see a clear
emergence of two sources, one which has a smaller correlation radius (see 
larger $q_{\text{out}}$) and the other which has a larger correlation radius. 
This
aspect has remained a recurring theme in results for photon intensity 
interferometry~\cite{dks} and has its origin in the emissions from quark
 and the hadronic phases respectively which have vastly differing source dimensions.
It is interesting that this feature noted in early exploratory studies
 has survived
the vast improvements in the rate calculations and the dynamics of evolution.
The emergence of the two source structure at larger $k_T$ is facilitated by the
decreasing contribution of the hadronic phase there.

 This feature is not expected for pion intensity correlations as they leave the
system mostly at the time of freeze-out only. 

The side-ward correlations show a decreasing radial dimension as $k_T$ 
increases,
as expected for a transversely expanding source, and look Gaussian in nature.
The longitudinal correlation function also shows a similar behaviour, as far
as the
variation of the correlation radius is concerned, however it is definitely not 
Gaussian in form. Thus neither the function (Eq.\ref{actual}) describes these
variations satisfactorily, nor does the procedure of taking $R_i=1/\tilde{q_i}$,
where $\tilde{q_i}$ denotes the momentum difference where $C-1$ has dropped by
a factor of $1/e$ compared to its value at $q_i=0$, can do a  full
justice to these distributions. 

In view of the above, we have numerically evaluated:
\begin{equation}
R_i^2=1/<q_i^2>;~~~ i={\text{ out, side, long}}
\end{equation}
where
\begin{equation}
<q_i^2>=\frac{\int \, dq_i \, q_i^2 \, (C-1)}{\int \, dq_i \, (C-1)}
\end{equation}
and plotted it as a function of $k_{1T}$ in Fig.~\ref{fig5}.

The $k_T$ variation of the $R_{\text{out}}$ reveals a very rich structure.
In order to fully appreciate the observations, we recall that
 the source of
photons has a dependence $\sim T^\nu \exp(-E/T)$ where $\nu \geq$ 2,
and $E$ is the energy of the photons in the rest frame of the fluid.  
This is very distinct from the corresponding dependence for pions, which 
varies as $\sim \exp(-E/T_f)$, where $T_f$ is the freeze-out temperature
and contributions accrue from locations along the freeze-out surface.
 The
``reclining chair'' behaviour seen in this variation is an outcome
 of competition
between a high temperature (limited to small radii and early times) 
and a large transverse flow velocity (predominant at large radii and late times)
in producing photons having moderate $k_T$. 
  Thus the large radial flow ensures
 that photons having  a large $k_T$ can also be produced at later times,
thus enhancing the so-called duration of the source measured as
 $R_{\text{out}}-R_{\text{side}}$. We shall see later that the the `seat' of
this ``reclining chair'' gets narrower for RHIC energies and vanishes at
LHC energies, as the increasing radial flow rapidly cools the system. A hint
of this behaviour is also present in the
momentum dependence of the side-ward correlation radius, which 
decreases almost linearly with $k_T$. The momentum dependence of the
 longitudinal
correlation radius is completely different from the $~1/\sqrt{m_T}$ 
expected for
pions~\cite{TC}. We see that $R_{\text{long}}$ for photons is almost inversely
 proportional
to the transverse momentum. We believe that these differences do arise 
due to the difference in the source function for photons and pions.

\subsection{One dimensional analysis and comparison to WA98 results}

Let us now look at our results for the one-dimensional analysis of the 
correlation function in terms of the invariant momentum difference 
$q_{\text{inv}}$
corresponding to the transverse momentum, and rapidity window used in the WA98
experiment~\cite{wa98_2}. Our results with all the kinematic cuts are shown in 
Fig.~\ref{fig6}. In order to simulate the probabilistic selection of photons,
we first generated a sufficiently large number of photons
according to the thermal distribution calculated by us earlier,
 in the transverse
momentum window of $k_T~\epsilon~[0.10,2.5]$ GeV, and 
(randomly) distributed them uniformly over
the azimuthal directions and the rapidity window corresponding to the 
experiment. Next we sampled pairs so that their average transverse momentum
$K_T$ was in the appropriate window. The correlation function was then 
calculated
using the expression Eq.(\ref{def}).
The results were then averaged by binning in $q_{\text{inv}}$.

We see that our results are described to a reasonable accuracy by the
form:
\begin{equation}
C=1+0.5\, a \, \exp\left[ -q_{\text{inv}}^2 R_{\text{inv}}^2/2\right ]
\label{fit}
\end{equation}
where $R_{\text{inv}}\approx$ 6.5 fm
 for $0.10 \leq k_T \leq 0.20$ GeV and 5.3 fm for the transverse momentum
window $0.20 \leq k_T \leq 0.30$ GeV. In order to compare our results with 
the numbers quoted the WA98 experiment~\cite{wa98_2} 
we need to multiply their results
with $\sqrt{2}$ as the fits discussed in that work do not include the factor
of 2 in the exponential used in the present work (Eq.\ref{one_1}).  Thus the 
corresponding experimental results for the $R_{\text{inv}}$ obtained 
by the WA98 experiment are 8.34 $\pm$ 1.7 fm and 8.63 $\pm$ 2.0 fm,
 respectively.
While our predictions are in reasonable agreement with the ``experimental''
findings (within the errors), the experimental results are on the larger side.
Before passing a judgment on our results, recall that that the values quoted by
the WA98 experiment are obtained by assigning a values of 7.85 and 7.25 fm 
for the $R_{\text{side}}$, 8.4 and 7.7 fm for the $R_{\text{long}}$ and 8.5 fm
for the $R_{\text{out}}$ based on the values obtained for
 pion interferometry~\cite{wa98_3}.

Our results for the $R_{\text{side}}$ are 4.93 and 4.85 fm,
 for the $R_{\text{long}}$
are 3.7 and 2.5 fm, and for the $R_{\text{out}}$ are
9.97 and 9.39 fm respectively at $k_T$ =0.15 and 0.25 GeV
respectively (see Fig.\ref{fig5}). Thus we feel that the results for
 the $R_{\text{inv}}$ as well as the results for the single photons
  at low $k_T$ quoted by the
WA98 experiment (based on the value of the parameter $a$ in the Eq.(\ref{fit}
above),
are model dependent and uncertain to the extent that they use correlation radii 
determined from pion interferometry. We also see that
 at least the side-ward and the 
longitudinal radii for photon correlations are much smaller than 
the corresponding
values for the pion interferometry. 

A full three dimensional determination of the correlation function 
will go a long way in getting reliable results and 
constraining the theoretical calculations discussed here.

\subsection{Results for RHIC and LHC energies}

We now present results for correlation functions for thermal
photons at RHIC and LHC energies (see Figs.\ref{fig7}--\ref{fig10}).

Looking at Fig.~\ref{fig7}, for central collisions of gold nuclei at
RHIC energies,  we note that the two-source aspect in the
out-ward correlation function becomes more clear, due to increased
contributions from the (smaller but stronger) quark-matter. Of-course, it is 
known~\cite{bms_hbt} that we shall have a large contribution of 
pre-equilibrium photons at larger $k_T$ and this trend should 
continue at LHC energies. This will further enhance the contribution
of the smaller-sized source, and the structure seen here, will
disappear.

The transverse momentum dependence of the source sizes( Fig.~\ref{fig8}), 
is similar in 
nature to what we see at  SPS energies, though the `seat' of the `reclining
chair' seen in the out-ward size gets narrower, as one would expect for a
more rapidly expanding source.
The longitudinal correlation length is again seen to decrease rapidly with
the increase in the transverse momentum.

 These trends continue at LHC energies (Fig.~\ref{fig9}) and the 
transverse momentum dependence of the source sizes (Fig.~\ref{fig10})
 becomes very pronounced.
The so-called `seat' in the outward correlation vanishes completely, and 
all the radii decrease roughly as $1/k_T^\alpha$, with $\alpha$ $\approx$
0.27 for the outward correlation, about 0.14 for the side-ward correlation
and about 0.94 for the longitudinal correlation. We note that at none of 
the energies, the longitudinal correlation function resembles a Gaussian,
as normally assumed in parametrization. This is perhaps related to the
emission of the photons from all the points in a longitudinally expanding
system. In fact, similar shape is seen for the longitudinal correlation
function for pions `escaping' from the system~\cite{grassi} and 
for pions produced from partonic cascades~\cite{klaus}.

\subsection{The life-time of the source}
The difference of the out-ward and side-ward correlation radii is often
associated with the life-time of the source. In order to see the evolution
of this parameter with the transverse momentum of the thermal photons we 
plot their difference as well as their ratios in Fig.~\ref{fig11}, for the
three systems studied in the present work. We see the `reclining chair'
behaviour for the  SPS energies, and a more rapid decrease as we go from
RHIC to LHC energies. Several interesting results emerge.

Firstly the difference $(R_{\text{out}}-R_{\text{side}})$ is nearly 
unaltered as we go from SPS to RHIC for photons having transverse momenta
less than 0.5 GeV. We have already noted that the `seat' of variation
is
caused  by the competition between cooling due to expansion and the
blue-shift of the momenta due to the transverse flow. The more rapid
decrease of the difference arises due to  a more rapid cooling and 
an increased
flow as the initial temperature rises. The same trend is seen for the 
ratios. An experimental confirmation or a refutation of theses 
predictions will be very valuable.

\section{Discussions and summary}
%%%%%%%%%%%%%%%%%%revision
Before summarizing, it may be of use to discuss some aspects which we
have overlooked in the present work.

It should  be of interest to examine the role of the reactions of the
type $ q+\bar{q} \rightarrow \gamma + \gamma$ or $ q + g \rightarrow q + \gamma
+ \gamma $, for the intensity interferometry of photons. 
Firstly, the quark matter contribution to single photons is already marginal
for low transverse momenta, which is the subject of our focus here. Then
we note such photon pairs (``diphotons'') will be produced at `a point' and 
not at different locations like a pair which comes from, say,
 $q +\bar{q} \rightarrow g + \gamma$ and $ q+ g \rightarrow q+ \gamma +\gamma$.
More-over their transverse momentum difference $\bf{q}_T \approx 2 k_T$, where 
$k_T$ is the transverse momentum of either photon. For calculations which
we can treat perturbatively, this difference is at least an order of magnitude
larger than the values of $\bf{q}_T$ where the correlation function has a
significant value. Of-course one could include them as a source of 
single photons. However they are suppressed by a factor $\alpha_s/\alpha$ 
compared to the $q+\bar{q} \rightarrow g+\gamma$, say.

Let us also discuss the role of direct (QCD) photons to the $k_T$ region
under consideration here and for which we have experimental data at
 SPS energies 
from the WA98 experiment. In addition to the results given here, calculations
have been reported by Dumitru et al.~\cite{dum}, where large values for
intrinsic momenta of partons are used to exhaust the WA98 data beyond 
$k_T \approx$ 3 GeV. The calculations exhaust only about 50\% of the
single photon yield at 2 GeV (as in the present work) and less than 20\% 
of the yield
at 1 GeV. Turbide et al.~\cite{simon} have extrapolated the pQCD results down
to zero transverse momenta (!) and find that the thermal photon yield is 
up to 4 orders of magnitude larger than the direct photon yield at lower
transverse momenta. Parton Cascade Model calculations for SPS energies
\cite{bms_phot}  
show the dominance of quark matter contributions beyond 3 GeV (including
photons which could be termed thermal, from the quark-matter) but a
 much smaller production from them at lower $k_T$. These considerations
should convince us that indeed thermal photons dominate in the region
of transverse momenta considered here.
%%%%%%%%%%%%%%%%%%%%%%%%%%%%%%%%%%%%%%%%%%%%%%%%%%%%%

We may also recall here the results of two works, which address similar
issues. Peressounko~\cite{ors} has used hydrodynamics calculations
with rates for the production of photons from quark matter
and hadronic matter, along with the photons from the decay of
pions to get $\lambda$ as well as the correlation radii, for direct
photons. While the $R_{\text{out}}$  and the $R_{\text{long}}$ 
vs. $k_T$  behaviour estimated by him are similar to ours, his work
shows a very unusual result: the $R_{\text{side}}$ actually 
increases, and that too substantially, with increase in $k_T$.
Considering that photons having large transverse momenta are produced
very early in the collision when the transverse expansion has not yet
set-in, this is hard to understand. Similarly, the decrease in the
values of the correlation radii, as $k_T$ decreases below 0.5 GeV
at SPS energies is also difficult to fathom. Other differences could be
attributed to the more complete rates for photon production used in
the present work.

The work of Renk~\cite{ors} is interesting in that it uses a
fire-ball description to model the collision at SPS and 
RHIC energies. The parameters of the
fireball are adjusted to reproduce the
pion spectra as well as the HBT radii for pions.
The differences seen in the present work then arise due to
the explicit appearance of a mixed phase here, during which the
speed of sound is zero and acceleration of the expansion is stalled.
Results of an investigation using different
equations of state, including one which does not
admit a mixed phase, will be published shortly.

In brief, the side-ward, out-ward, and longitudinal correlation
functions for intensity interferometry of thermal photons at SPS,
RHIC, and LHC energies are calculated. The transverse momentum
dependence of these radii are very different from the corresponding
results for pions, which are expected to decrease as $1/\sqrt{m_T}$
for all the components. The longitudinal correlations for the
three energies are quite similar and may be indicative of 
boost-invariance of the flow assumed in the work. The ratio
$R_{\text{out}}/R_{\text{side}}$ at LHC decreases rapidly and 
approaches unity, due to increase in $R_{\text{side}}$ due to expansion
and decrease in $R_{\text{out}}$ due to rapid cooling.

As the results are free from distortions due to final state 
interactions and uncertainties about the production-vertexes,
or the production mechanism of the particles under investigation,
a confirmation or refutation of the findings will be very
valuable.

It is also pointed out that the one dimensional analysis in
terms of the variable $Q_{\text{inv}}$ may have only a limited use.

\begin{acknowledgments}  
Discussions with Terry Awes, S. A. Bass, C. Gale,  M. G. Mustafa,
and D. Peressounko are gratefully acknowledged.
\end{acknowledgments}

\end{document}